\documentclass[a4paper,12pt]{article}
\usepackage[utf8x]{inputenc}
\usepackage{fancyhdr}
\usepackage{graphicx}
\usepackage{amsmath}
\usepackage{amssymb}
\usepackage{hyperref}
\usepackage[all]{hypcap}
\usepackage{pdfpages}
\usepackage{amsthm}
\usepackage{caption}
\usepackage{subcaption}
\usepackage[margin=1in]{geometry}
\title{A bi-convex optimization problem to compute Nash equilibrium in n-player games and an algorithm.}
\author{Vinayaka G. Yaji and Shalabh Bhatnagar\\ \it{Department of Computer Science and Automation}\\ \it{Indian Institute of Science, Bangalore, India.}\\email: vgyaji@gmail.com, shalabh@csa.iisc.ernet.in}

\begin{document}

\maketitle

\begin{abstract}
In this paper we present optimization problems with biconvex objective function and linear constraints such that the set of global minima of the optimization problems is the same as the set of Nash eqilibria of a n-player general-sum normal form game. We further show that the objective function is an invex function and consider a projected gradient descent algorithm. We prove that the projected gradient descent scheme converges to a partial optimum of the objective function. We also present simulation results on certain test cases showing convergence to a Nash equilibrium strategy.
\end{abstract}

\section{Introduction}
A general theory of games first introduced in \cite{vnom} has found several applications in the field of economics and engineering. A solution concept or a notion of equilibrium was proposed by Nash (known as Nash equilibrium) in \cite{nash} and was shown to exist in every finite normal-form game. Further generalizations of Nash equilibrium such as correlated equilibrium and coarse correlated equilibrium were also introduced and studied. It is well known that for every game the set correlated and coarse-correlated equilibria are convex subsets of the strategy space. But in general the set of Nash equilibria is not convex.
A number of methods have been proposed to compute a Nash equilibium strategy. Lemke-Howson's algorithm for bi-matrix games\cite{lemke}, global newton method\cite{govind}, homotopy based methods\cite{herrings} are some of the few methods to compute a Nash equilibrium strategy.

For a general n-player game, the associated optimization problem is non-linear and non-convex and hence is difficult to solve. It is known that the problem of computing nash equilibria in bi-matrix games is a linear complementarity problem and for the general n-player scenario it is a non-linear complementarity problem. 
Linear complimentarity problems (the ones arising from games) can be solved using Lemke-Howson's method, while non-linear complimentarity problems are in general hard to solve and require some sufficient conditions to be imposed on the problem to solve them which is not satisfied by every game. 

In this paper we present optimization problems with biconvex objective function and linear constraints such that the set of global minima of the optimization problems is the same as the set of Nash eqilibria of a n-player general-sum normal form game. Global optimization algorithms exist that can compute the global minima of such optimization problems\cite{floudas}. The main idea in the formulation of these optimization problems is the fact that correlated or coarse-correlated equilibrium which are product of individual player's strategy is a Nash equilibrium. 
We further show that the objective function is an invex function i.e. the set of stationary points is the same as the set of global minima. We also consider a projected gradient descent scheme and prove that is converges to a partial optimum of the objective function. 

The remainder of this paper is organised as follows: In section 2, necessary definitions and notations are stated. In section 3, functions with required properties are defined. In section 4, properties of the functions defined in section 2 are proved. In section 5, optimization problems are presented. In section 6, the projected gradient descent algorithm is stated and convergence analysis is performed. In section 7, simulation results of the projected gradient descent algorithm on certain test cases are presented. In section 8, we summarize and present directions for future research.

\section{Definitions and notations.} 
In this section we shall state definitions, introduce variables and notations used later in this paper. 

A normal form game (or simply a game) ($\Gamma$) is defined by tuple $\Gamma=<I,\ \{A^i\}_{i\in I},\ \{u^{i}\}_{i\in I}>$ where, $I$ denotes the set of players ($I=\{1,\dots,N\}$), $\forall i\in I,\ A^i$ denotes the set of actions of player $i$ ($A^{i}=\{a^i_j\ :\ 1\leq j\leq m_i\}$). Let $A=\times_{i \in I}A^i$ and $\forall i \in I,\ u^i:A\rightarrow\mathbb{R}$ denotes the utility function of player $i$.

For every $i\in I$, $\Sigma^{i}$ denotes the set of probability distributions on $A^{i}$. $\Sigma^{i}$ is identified by the probability simplex $\bigtriangleup^{m_i}\subseteq\mathbb{R}^{m_i}$. $\pi^{i}=(\pi^{i}(a^i_1),\dots,\pi^{i}(a^i_{m_i}))$ denotes a generic element of $\Sigma^{i}$.
Let $\pi=<\pi^1,\dots,\pi^{m_i}>$ which is identified as a vector in $\times_{i \in I}\bigtriangleup^{m_i}\subseteq\mathbb{R}^{M_1}$ where $M_1=\sum\limits_{i\in I}m_i$. Let $\Sigma=\times_{i\in I}\Sigma^{i}$. 

Let $\Sigma_{C}$ denote the set of probability distributions on $A$. $\Sigma_{C}$ is identified by the probability simplex $\bigtriangleup^{M_2}\subseteq\mathbb{R}^{M_2}$ where $M_2=\prod\limits_{i\in I}m_i$. 
$p=(p(a) : a\in A)$ denotes a generic element in $\Sigma_{C}$. 

For every $i\in I$, $A^{-i}=\times_{\{k\in I,\ k\neq i\}}A^{k}$ and $a^{-i}$ denotes a generic element in $A^{-i}$. Similarly, this can be extended to more than one player. 
$\forall i\in I,\ \forall a^{-i}=(a^k_{j_k}\ :\ k\in I,\ k\neq i,\ a^k_{j_k}\in A^{k}) \in A^{-i},\ \forall a^i_{j_i} \in A^i,\ (a^i_{j_i},a^{-i})=(a^1_{j_1},\dots,\ a^{i-1}_{j_{i-1}},\ a^{i}_{j_i},\ a^{i+1}_{j_{i+1}},\dots,\ a^N_{j_N}) \in A$.
 Similarly define $\forall i\in I,\ \Sigma^{-i}=\times_{\{k\in I,\ k\neq i\}}\Sigma^{k}$ and $\pi^{-i}$ denote a generic element in $\Sigma^{-i}$. $\forall i\in I,\ \forall \pi^{-i}=(\pi^k\ :\ k\in I,\ k\neq i,\ \pi^k\in \Sigma^{k}) \in \Sigma^{-i},\ \forall \pi^i \in \Sigma^i,\ (\pi^i,\pi^{-i})=(\pi^1,\dots,\ \pi^{i-1},\ \pi^{i},\ \pi^{i+1},\dots,\ \pi^N) \in \Sigma$.

For every $i\in I$, $u^i(\pi)=\sum\limits_{a\in A}u^i(a)\prod\limits_{i\in I}\pi^{i}(a^i_{j_i})$ where $a=(a^i_{j_i}\ :\ i\in I)$. For every $i\in I$, $\forall a^i_j\in A^i,\ \forall \pi^{-i}\in\Sigma^{-i},\ u^i(a^i_j,\pi^{-i})=\sum\limits_{a^{-i}\in A^{-i}}u^{i}(a^i_j,a^{-i})\prod\limits_{k\in I,\ k\neq i}\pi^k(a^k_{j_k})$ where $a^{-i}=(a^k_{j_k}\ :\ k\in I,\ k\neq i)$. 

For every $i\in I$, $u^i(p)=\sum\limits_{a\in A}u^i(a)p(a)$ and $\forall a^{i}_{j_i}\in A^i,\ u^i(a^i_{j_i},p^{-i})=\sum\limits_{a^{-i}\in A^{-i}}u^i(a^i_{j_i},a^{-i})\sum\limits_{j=1}^{m_i}p(a^i_j,a^{-i})$. Similarly define $\forall i\in I,\ \forall \pi^{i}\in \Sigma^{i},\ \forall p\in \Sigma_{C},\ u^i(\pi^{i},p^{-i})=\sum\limits_{j=1}^{m_i}u^i(a^i_j,p^{-i})\pi^i(a^i_j)$.  

$\pi\in \Sigma$ is said to be a \bf{Nash equilibrium }\rm strategy of the game $\Gamma$ (or just N.E.) if $\forall i\in I,\ \forall a^i_j\in A^{i},\ u^i(a^i_j,\pi^{-i})-u^i(\pi)\leq0$. Let $NE(\Gamma)$ denote the set of Nash equilibria strategies of game $\Gamma$.

$p\in \Sigma_{C}$  is said to be a \bf{correlated equilibrium }\rm strategy of the game $\Gamma$ (or just C.E.) if $\forall i\in I,\ \forall a^i_j,a^i_{j^{'}}\in A^{i},\ \sum\limits_{a^{-i}\in A^{-i}}(u^i(a^i_{j^{'}},a^{-i})-u^i(a^i_{j},a^{-i}))p(a^i_j,a^{-i})\leq0$. Let $CE(\Gamma)$ denote the set of correlated equilibria of the game $\Gamma$.

$p\in \Sigma_{C}$ is said to be a \bf{coarse correlated equilibrium }\rm strategy of the game $\Gamma$ (or just C.C.E.) if $\forall i\in I,\ \forall a^i_j\in A^i,\ u(a^{i}_{j},p^{-i})-u^{i}(p)\leq0$. Let $CCE(\Gamma)$ denote the set of coarse correlated equilibria of the game $\Gamma$.

Define $P : \Sigma \rightarrow \Sigma_{C}$, s.t. , $\forall \ \pi \in \Sigma,\ \forall \ a \in A,\ P(\pi)(a)=\prod_{\substack{i \in I}}\pi^{i}(a^i_{j_i})$ where $a=(a^{i}_{j_i}:i\in I)$.
Let the graph of the function $P$ be $\mathcal{G}(P):=\{(\pi,p)\in \Sigma \times \Sigma_{C} : p=P(\pi)\}$. In the following lemma we summarize the relationship between the various equilibrium concepts defined.\newline\newline
\bf{Lemma 2.1: }\rm Given a game $\Gamma$. The following hold.  
\begin{itemize}
 \item [(1)] $P(NE(\Gamma))\subseteq CE(\Gamma)\subseteq CCE(\Gamma)$.
 \item [(2)] $p\in CE(\Gamma),\ \exists \pi\in \Sigma,\ s.t.,\ p=P(\pi),\ then,\ \pi\in NE(\Gamma)$.
 \item [(3)] $p\in CCE(\Gamma),\ \exists \pi\in \Sigma,\ s.t.,\ p=P(\pi),\ then,\ \pi\in NE(\Gamma)$.
\end{itemize}

The results in lemma follow directly from definitions.

$(\pi,p)\in \Sigma\times\Sigma_{C}$ is a \bf{Nash equilibrium profile }\rm of game $\Gamma$ if $\pi$ is a Nash equilibrium strategy of game $\Gamma$ and $p=P(\pi)$. 

Let $A_1$ and $A_2$ be two convex subsets of $\mathbb{R}^{n_1}$ and $\mathbb{R}^{n_2}$ respectively. A function $g:A_1\times A_2\rightarrow\mathbb{R}$ is said to be a \bf{biconvex function }\rm if $\forall x\in A_1,\ g(x,\cdot):A_2\rightarrow\mathbb{R}$ is a convex function and $\forall y\in A_2,\ g(\cdot,y):A_1\rightarrow\mathbb{R}$ is a convex function. $(x^{*},y^{*})\in A_1\times A_2$ is a \bf{partial optimum }\rm of a biconvex function $g$ if $\forall x\in A_1,\ g(x^*,y^*)\leq g(x,y^*)$ and $\forall y\in A_2,\ g(x^*,y^*)\leq g(x^*,y)$. For a detailed study of biconvex functions see \cite{gorksi}.

Let $\mathcal{F}$ be a subset of $\mathbb{R}^n$ and $g:\mathcal{F}\rightarrow\mathbb{R}$. $x^{*}\in \mathcal{F}$ is said to the global optimum of the optimization problem $\min_{x}\ g(x),\ subject\ to,\ x\in \mathcal{F}$, if, $\forall x\in \mathcal{F},\ g(x^*)\leq g(x)$.

\section{Objective functions.}
 In this section we shall define functions whose set of zeros is the same as the set of Nash equilibria of the game $\Gamma$. 
 The following theorem gives a necessary and sufficient condition for $(\pi,p)\in \Sigma \times \Sigma_{C}$ to be in $\mathcal{G}(P)$.\newline\newline
 \bf{Theorem 3.1: }\rm
  Given $(\pi,p)\in \Sigma \times \Sigma_{C}$. Then,
  $(\pi,p) \in \mathcal{G}(P)$ iff $\forall i \in I,\ \forall a \in A,\ p(a)-\pi^i(a^i_{j_i})\sum_{\substack{j=1}}^{m_i}p(a^i_j,a^{-i}) = 0,\ where\ a=(a^{i}_{j_i},a^{-i})$.\newline
 \bf{Proof : }\rm [$\Rightarrow$] Assume $(\pi,p) \in \mathcal{G}(P)$. Fix $i \in I,\ a \in A\ (where\ a=(a^k_{j_k}:k \in I))$. Then$\ p(a)=\prod_{\substack{k \in I}}\pi^{k}(a^k_{j_k})$ and $\sum_{\substack{j=1}}^{m_i}p(a^i_j,a^{-i})=
 \sum_{\substack{j=1}}^{m_i}\pi^i(a^i_j)\prod_{\substack{k \in I \\ k\neq i}}\pi^{k}(a^k_{j_k})=\\\prod_{\substack{k \in I \\ k\neq i}}\pi^{k}(a^k_{j_k})\sum_{\substack{j=1}}^{m_i}\pi^i(a^i_j)=\prod_{\substack{k \in I \\ k\neq i}}\pi^{k}(a^k_{j_k})$. 
 Therefore $\ p(a)-\pi^i(a^i_{j_i})\sum_{\substack{j=1}}^{m_i}p(a^i_j,a^{-i})\ = \ \prod_{\substack{k \in I}}\pi^{k}(a^k_{j_k})-\pi^i(a^i_{j_i})\prod_{\substack{k \in I \\ k\neq i}}\pi^{k}(a^k_{j_k})\ = \ 0$. 
 Since $i \in I,\ a\in A$ are arbitrary, $\ p(a)-\pi^i(a^i_{j_i})\sum_{\substack{j=1}}^{m_i}p(a^i_j,a^{-i})=0$, $\forall i \in I\ and\ \forall a \in A$.\newline
 [$\Leftarrow$] Fix $a_{*}\in A\ where\ a_{*}=(a^i_{j^{*}_i}:i\in I)=(a^i_{j^{*}_i}:1\leq i \leq N)$. 
 From data, we know that $\forall a^{-1}\in A^{-1},\ p(a^1_{j^{*}_1},a^{-1})=\pi^{1}(a^{1}_{j^{*}_1})\sum_{j_1=1}^{m_1}p(a^1_{j_1},a^{-1})$.
 Using the above, we get, $\forall a^{-1,2}\in A^{-1,2},\ \sum_{j_2=1}^{m_2}p(a^1_{j^{*}_1},a^2{j_2},a^{-1,2})=\pi^1(a^1_{j^{*}_1})\sum_{j_2=1}^{m_2}\sum_{j_1=1}^{m_1}p(a^1_{j_1},a^2_{j_2},a^{-1,2})$.
 From data, we also know that $\forall a^{-1,2}\in A^{-1,2},\ p(a^1_{j^{*}_1},a^2_{j^{*}_2},a^{-1,2})=\pi^{2}(a^{2}_{j^{*}_2})\sum_{j_2=1}^{m_2}p(a^{1}_{j^{*}_1},a^2_{j_2},a^{-1,2})$.
 Therefore by substituting for the sum, we get, $\forall a^{-1,2}\in A^{-1,2},\ p(a^1_{j^{*}_1},a^2_{j^{*}_2},a^{-1,2})=\\\pi^{2}(a^{2}_{j^{*}_2})\pi^1(a^1_{j^{*}_1})\sum_{j_2=1}^{m_2}\sum_{j_1=1}^{m_1}p(a^1_{j_1},a^2_{j_2},a^{-1,2})$. 
 Similarly repeating the above procedure for actions of the third player we get, $\forall a^{-1,2,3}\in A^{-1,2,3},\ p(a^1_{j^{*}_1},a^2_{j^{*}_2},a^3_{j^{*}_3},a^{-1,2,3})=\\\pi^{3}(a^{3}_{j^{*}_3})\pi^{2}(a^{2}_{j^{*}_2})\pi^1(a^1_{j^{*}_1})\sum_{j_3=1}^{m_3}\sum_{j_2=1}^{m_2}\sum_{j_1=1}^{m_1}p(a^1_{j_1},a^2_{j_2},a^3_{j_3},a^{-1,2,3})$.   
 Proceeding all the way upto player $N$ we get, $p(a_{*})=(\prod_{i \in I}\pi^{i}(a^i_{j^{*}_i}))(\sum_{j_N=1}^{m_N}...\sum_{j_1=1}^{m_1}p(a^{1}_{j_1},...,a^{N}_{j_N}))$.
 Since $p\in \Sigma_{C}$, we know that $\sum_{a\in A}p(a)=\sum_{j_N=1}^{m_N}...\sum_{j_1=1}^{m_1}p(a^{1}_{j_1},...,a^{N}_{j_N})=1$.
 Therefore, $p(a_{*})=\prod_{i \in I}\pi^{i}(a^i_{j^{*}_i})$. Since $a_{*} \in A$ is arbitrary, $p(a_{*})=\prod_{i \in I}\pi^{i}(a^i_{j^{*}_i})\ \forall a_{*} \in A$.\hfill $\blacksquare$\newline
 
 Using the above theorem we now define a non-negative function on $\Sigma \times \Sigma_{C}$ such that the function takes the value zero on $\mathcal{G}(P)$ and is positive on $\mathcal{G}(P)^C$.
 
 Let $f:\Sigma \times \Sigma_{C}\rightarrow[0,\infty)$ such that, $\forall (\pi,p)\in \Sigma \times \Sigma_{C},\ f(\pi,p)=\sum\limits_{i \in I}\sum\limits_{\substack{a\in A \\ a=(a^i_{j_i},a^{-i})}}(p(a)-\pi^i(a^i_{j_i})\sum_{\substack{j=1}}^{m_i}p(a^i_j,a^{-i}))^2$.\newline\newline 
 \bf{Corollary 3.1: }\rm 
 Given $(\pi,p)\in \Sigma \times \Sigma_{C}$. Then,
 $f(\pi,p)=0$ iff $(\pi,p) \in \mathcal{G}(P)$. \newline
 
 From the definitions of coarse-correlated equilibrium and correlated equilibrium we now define the following non-negative functions on $\Sigma_{C}$ such that they take the value zero on the set of 
 coarse-correlated equilibria ($CCE(\Gamma)$) and correlated equilibria ($CE(\Gamma)$) respectively.
 
 Let $C_1:\Sigma_{C}\rightarrow[0,\infty)$, such that, $\forall p\in \Sigma_{C},\ C_1(p)=\sum\limits_{i \in I}\sum\limits_{j=1}^{m_i}(max\{u(a^{i}_{j},p^{-i})-u^{i}(p),0\})^2$
 and $C_2:\Sigma_{C}\rightarrow[0,\infty)$, such that, $\forall p\in \Sigma_{C},\\ C_2(p)=\sum\limits_{i \in I}\sum\limits_{j=1}^{m_i}\sum\limits_{j^{'}=1}^{m_i}(max\{\sum\limits_{a^{-i}\in A^{-i}}(u^i(a^i_{j^{'}},a^{-i})-u^i(a^i_{j},a^{-i}))p(a^i_j,a^{-i}),0\}^2$.\newline\newline
 \bf{Lemma 3.1: }\rm Given $p \in \Sigma_{C}$.
 \begin{itemize}
  \item $C_1(p)=0$ iff $p\in CCE(\Gamma)$.
  \item $C_2(p)=0$ iff $p\in CE(\Gamma)$.
 \end{itemize}
\bf{Proof : }\rm Follows directly from the definitions of correlated equilibrium and coarse correlated equilibrium in section 2.\hfill $\blacksquare$\newline

Let $B:\Sigma\times\Sigma_{C}\rightarrow[0,\infty)$ s.t. $\forall (\pi,p)\in\Sigma\times\Sigma_{C},\ B(\pi,p)=\sum\limits_{i\in I}\sum\limits_{j=1}^{m_i}(max\{u(a^{i}_{j},p^{-i})-u^{i}(\pi^{i},p^{-i}),0\})^2$.
 The idea is that when $(\pi,p)\in\mathcal{G}(P)$ and $B(\pi,p)=0$, then, $\forall i\in I,\pi^i$ is a best response to $\pi^{-i}$.\newline\newline 
\bf{Lemma 3.2: }\rm Given $(\pi,p)\in\mathcal{G}(P)$. $B(\pi,p)=0$ iff $\pi$ is a Nash equilibrium.\newline
\bf{Proof : }\rm [$\Rightarrow$]Since $B(\pi,p)=0$, we have, $\forall i\in I,\ \forall j\in\{1,\dots,m_i\},\ max\{u(a^{i}_{j},p^{-i})-u^{i}(\pi^{i},p^{-i}),0\}=0$. Hence $\forall i\in I,\ \forall j\in\{1,\dots,m_i\},\ u(a^{i}_{j},p^{-i})-u^{i}(\pi^{i},p^{-i})\leq0$. Since $(\pi,p)\in \mathcal{G}(P)$, $u^i(a^i_j,p^{-i})=u^i(a^i_j,\pi^{-i})$ and $u^i(\pi^{i},p^{-i})=u^i(\pi^{i},\pi^{-i})$.
Therefore, $\forall i\in I,\ \forall j\in\{1,\dots,m_i\},\ u(a^{i}_{j},\pi^{-i})-u^{i}(\pi^{i},\pi^{-i})\leq0$, which by definition of a Nash equilibrium strategy in section 2, implies $\pi$ is Nash equilibrium.

[$\Leftarrow$] Since $\pi$ is a Nash equilibrium, we have, $\forall i\in I,\ \forall j\in\{1,\dots,m_i\},\ u(a^{i}_{j},\pi^{-i})-u^{i}(\pi^{i},\pi^{-i})\leq0$. Since $(\pi,p)\in \mathcal{G}(P)$, $u^i(a^i_j,p^{-i})=u^i(a^i_j,\pi^{-i})$ and $u^i(\pi^{i},p^{-i})=u^i(\pi^{i},\pi^{-i})$. Therefore, $\forall i\in I,\ \forall j\in\{1,\dots,m_i\},\ u(a^{i}_{j},p^{-i})-u^{i}(\pi^{i},p^{-i})\leq0$, which further implies, $\forall i\in I,\ \forall j\in\{1,\dots,m_i\},\ max\{u(a^{i}_{j},p^{-i})-u^{i}(\pi^{i},p^{-i}),0\}=0$. Thus $B(\pi,p)=0$.\hfill $\blacksquare$\newline

We now characterise the set of nash equilibria of a game ($\Gamma$) using the functions $f,\ B,\ C_1$ and $C_2$.\newline\newline 
\bf{Theorem 3.2: }\rm Given $(\pi,p)\in \Sigma\times\Sigma_{C}$.
\begin{itemize}
 \item[(1)] $(\pi,p)$ is a Nash equilibrium profile iff $f(\pi,p)+C_1(p)=0$.
 \item[(2)] $(\pi,p)$ is a Nash equilibrium profile iff $f(\pi,p)+C_2(p)=0$.
 \item[(3)] $(\pi,p)$ is a Nash equilibrium profile iff $f(\pi,p)+B(\pi,p)=0$.
\end{itemize}
\bf{Proof : }\rm First we shall prove (1).[$\Rightarrow$] Assume $(\pi,p)$ is a Nash equilibrium. Then, by definition of Nash equilibrium profile in section 2, $\pi$ is a N.E. and 
$p=P(\pi)$. By lemma 2.1, since $\pi$ is a N.E. $P(\pi)=p\in CCE(\Gamma)$ and since $p=P(\pi)$, $(\pi,p)\in \mathcal{G}(P)$. Thus $f(\pi,p)=0$ and $C_1(p)=0$ by
Theorem 3.1 and Lemma 3.1 respectively. Therefore $f(\pi,p)+C_1(p)=0$.\newline
[$\Leftarrow$] Assume $f(\pi,p)+C_1(p)=0$. Since both $f$ and $C_1$ are non-negative, $f(\pi,p)=0$ and $C_1(p)=0$. By Theorem 3.1, $f(\pi,p)=0$
will imply $(\pi,p)\in \mathcal{G}(P)$ and by Lemma 3.1 $C_1(p)=0$ will imply $p\in CCE(\Gamma)$. Since $p\in CCE(p)$ and $p=P(\pi)$, from Lemma 2.1,
we have that $\pi$ is a N.E. Thus $(\pi,p)$ is a Nash equilibrium.

Proof of (2) is similar  to that of (1) and the proof of (3) follows from Lemma 3.2 and corollary 3.1.\hfill $\blacksquare$\newline

\section{Properties of the objective functions.}
In this section we shall prove certain properties of the functions constructed in section \bf{no}\rm. First, we shall prove that $f$ is biconvex and that $C_1$ and $C_2$ are convex.\newline\newline 
\bf{Lemma 4.1: }\rm $f$ is a biconvex function i.e. $\forall \pi \in \Sigma,\ f(\pi,.):\Sigma_{C}\rightarrow[0,\infty)$ is convex and $\forall p\in \Sigma_{C},\ f(.,p):\Sigma \rightarrow[0,\infty)$ is convex.\newline
\bf{Proof : }\rm $\forall i\in I,\ \forall a\in A$ where $a=(a^{i}_{j_i}:i\in I)$, $p(a)-\pi^i(a^i_{j_i})\sum_{\substack{j=1}}^{m_i}p(a^i_j,a^{-i})$ is a linear function of $p\in \Sigma_{C}$ and an affine function of $\pi\in\Sigma$. By proposition 1.1.4 in \cite{bert}, 
$(p(a)-\pi^i(a^i_{j_i})\sum_{\substack{j=1}}^{m_i}p(a^i_j,a^{-i}))^2$ is convex in $p\in \Sigma_{C}$ and $\pi\in \Sigma$ with the other fixed. Since sum of convex functions is convex, $f(\pi,p)=\sum\limits_{i \in I}\sum\limits_{\substack{a\in A \\ a=(a^i_{j_i},a^{-i})}}(p(a)-\pi^i(a^i_{j_i})\sum_{\substack{j=1}}^{m_i}p(a^i_j,a^{-i}))^2$ is convex in $p$ for every fixed $\pi\in \Sigma$ and is convex in $\pi$ for every fixed $p\in\Sigma_{C}$.\hfill $\blacksquare$\newline\newline 
\bf{Lemma 4.2: }\rm $C_{1}$ and $C_{2}$ are convex functions of $p\in \Sigma_{C}$.\newline
\bf{Proof : }\rm First we shall show $C_1$  is convex. $\forall i\in I,\ \forall j\in \{ 1,\dots,m_{i}\}$, $u(a^{i}_{j},p^{-i})-u^{i}(p)$ is linear in $p\in \Sigma_{C}$. Since supremum of convex functions 
is convex, we have,  $\forall i\in I,\ \forall j\in \{ 1,\dots,m_{i}\}$, $ max\{u(a^{i}_{j},p^{-i})-u^{i}(p),0\}$. Since composition of nondecreasing function and convex function is convex, $\forall i\in I,\ \forall j\in \{ 1,\dots,m_{i}\}$, $ max\{u(a^{i}_{j},p^{-i})-u^{i}(p),0\}^2$, is convex. Therefore, 
$C_1(p)=\sum\limits_{i \in I}\sum\limits_{j=1}^{m_i}(max\{u(a^{i}_{j},p^{-i})-u^{i}(p),0\})^2$ is a convex function. 

Similarly we can show that $C_{2}$ is also a convex function.\hfill $\blacksquare$\newline

It is easy to show $f,\ C_{1}$ and $C_{2}$ are continuously differentiable on an open set containing their respective domains (for a similar proof refer \cite{mckelvy}).
Let $\nabla f(\pi,p)=[\nabla_{\pi}f(\pi,p)^T\ \nabla_{p}f(\pi,p)^T]^T$, where $\nabla_{\pi}f(\pi,p)=(\frac{\partial f(\pi,p)}{\partial \pi^{i}(a^{i}_j)}:\ i\in I,\ 1\leq j\leq m_{i})$ and
$\nabla_{p}f(\pi,p)=(\frac{\partial f(\pi,p)}{\partial p(a)}:\ a\in A)$. For every $k\in I,\ \forall j\in\{1,\dots,m_{k}\}$,

\begin{align*}
\frac{\partial f(\pi,p)}{\partial \pi^{k}(a^{k}_{j})}&=\sum\limits_{i \in I}\sum\limits_{\substack{a\in A \\ a=(a^i_{j_i},a^{-i})}}\frac{\partial}{\partial \pi^{k}(a^{k}_j)} (p(a)-\pi^i(a^i_{j_i})\sum_{\substack{\hat{j}=1}}^{m_i}p(a^i_{\hat{j}},a^{-i}))^2\\
&=\sum\limits_{\substack{a\in A \\ a=(a^k_{j_k},a^{-k})}}\frac{\partial}{\partial \pi^{k}(a^{k}_j)} (p(a)-\pi^i(a^k_{j_k})\sum_{\substack{\hat{j}=1}}^{m_k}p(a^k_{\hat{j}},a^{-k}))^2\\
&=-2[\sum\limits_{\substack{a^{-k}\in A^{-k}}}(p(a^k_j,a^{-k})-\pi^i(a^k_{j_k})\sum_{\substack{\hat{j}=1}}^{m_k}p(a^k_{\hat{j}},a^{-k}))\sum_{\substack{\hat{j}=1}}^{m_k}p(a^k_{\hat{j}},a^{-k})]
\end{align*}

So as to compute $\nabla_{p}f(\pi,p)$, we shall write $f(\pi,p)=\sum\limits_{i \in I}\sum\limits_{\substack{a\in A \\ a=(a^i_{j_i},a^{-i})}} (h^{i,a}(\pi)^Tp)^2$, where $h^{i,a}(\pi)\in\mathbb{R}^{M_2}$ s.t. 
$\forall i \in I,\ \forall a \in A,\ p(a)-\pi^i(a^i_{j_i})\sum_{\substack{j=1}}^{m_i}p(a^i_j,a^{-i}) = h^{i,a}(\pi)^Tp$ (which is possible since $p(a)-\pi^i(a^i_{j_i})\sum_{\substack{j=1}}^{m_i}p(a^i_j,a^{-i})$ is linear in $p$). Therefore,
\begin{align*}
\nabla_{p}f(\pi,p)&=\sum\limits_{i \in I}\sum\limits_{\substack{a\in A \\ a=(a^i_{j_i},a^{-i})}} \nabla_{p}(h^{i,a}(\pi)^Tp)^2\\
&=2\sum\limits_{i \in I}\sum\limits_{\substack{a\in A \\ a=(a^i_{j_i},a^{-i})}}(h^{i,a}(\pi)^Tp)h^{i,a}(\pi)
\end{align*}

The following lemma says that set of partial optima of $f$, the set of stationary points of $f$ and the set of global minima of $f$ are all the same.\newline\newline
\bf{Lemma 4.3: }\rm Given $(\pi^{*},p^{*})\in \Sigma\times\Sigma_{C}$. Then the following are equivalent.
\begin{itemize}
 \item [(1)] $(\pi^{*},p^{*})$ is a partial optimum of $f$.
 \item [(2)] $(\pi^{*},p^{*})$ is s.t. $f(\pi,p)=0$.
 \item [(3)] $(\pi^{*},p^{*})$ is s.t. $\nabla f(\pi^{*},p^{*})=0$.
\end{itemize}
\bf{Proof : }\rm[$(1)\Rightarrow(2)$]. Since $(\pi^{*},p^{*})$ is a partial optimum of $f$, $\forall p\in\Sigma_{C},\ f(\pi^{*},p^{*})\leq f(\pi^{*},p)$. Hence, 
$0\leq f(\pi^{*},p^{*})\leq f(\pi^{*},P(\pi^{*}))=0$. Therefore, $f(\pi^{*},p^{*})=0$.

[$(2)\Rightarrow(3)$]. Since $f(\pi,p)=0$, $\forall i \in I,\ \forall a \in A,\ p(a)-\pi^i(a^i_{j_i})\sum_{\substack{j=1}}^{m_i}p(a^i_j,a^{-i}) = 0,\ where\ a=(a^{i}_{j_i},a^{-i})$.
Substituting the above in the expression of $\nabla_{\pi}f(\pi,p)$ and $\nabla_{p}f(\pi,p)$ we get, $\nabla f(\pi^{*},p^{*})=0$. 

[$(3)\Rightarrow(1)$]. Since $f$ is biconvex (from Lemma 4.1), $f(.,p^{*})$ and $f(\pi^{*},.)$ are convex functions. From proposition 1.1.7 in \cite{bert}, we get,
 $\forall \pi\in\Sigma,\ f(\pi,p^{*})\geq f(\pi^{*},p^{*})+\nabla_{\pi}f(\pi^{*},p^{*})^T(\pi-\pi^{*})$ and $\forall p\in\Sigma_{C},\ f(\pi^{*},p)\geq f(\pi^{*},p^{*})+\nabla_{p}f(\pi^{*},p^{*})^T(p-p^{*})$.
 Substituting $\nabla f(\pi^{*},p^{*})=[\nabla_{\pi}f(\pi^{*},p^{*})^T\ \nabla_{p}f(\pi^{*},p^{*})^T]^T=0$, will give, $\forall \pi\in\Sigma,\ f(\pi,p^{*})\geq f(\pi^{*},p^{*})$ and $\forall p\in\Sigma_{C},\ f(\pi^{*},p)\geq f(\pi^{*},p^{*})$.
 Thus, $(\pi^{*},p^{*})$ is a partial optimum of $f$.\hfill $\blacksquare$\newline
 
 So as to compute $\nabla_{p}C_1(p)$, we shall write $C_1(p)=\sum\limits_{i \in I}\sum\limits_{j=1}^{m_i}(max\{(g^{i,j})^Tp,0\})^2$ where $\forall i\in I,\ \forall j\in\{1,\dots,m_i\},\ g^{i,j}\in\mathbb{R}^{M_2},\ s.t.,\ (g^{i,j})^Tp=u(a^{i}_{j},p^{-i})-u^{i}(p)$ (which is possible since $u(a^{i}_{j},p^{-i})-u^{i}(p)$ is linear in $p$).
 Then $\nabla_{p}C_1(p)=2\sum\limits_{i \in I}\sum\limits_{j=1}^{m_i}(max\{(g^{i,j})^Tp,0\})g^{i,j}$.
  
 The following lemma says that the set of global minima of $C_1$ and the set of stationary points of $C_1$ are the same.\newline\newline
 \bf{Lemma 4.4: }\rm Given $p^{*}\in\Sigma_{C}$. $C_1(p^{*})=0$ iff $\nabla_{p}C_1(p^{*})=0$.\newline
 \bf{Proof : }\rm Follows directly from the expression of the gradient and the convexity of $C_1$.\hfill $\blacksquare$\newline
 
 A similar result can be derived for $C_2$. In what follows in this paper results proved for $C_1$ can be extended to $C_2$ as well. 
 
 In theorem 3.2 we showed that the set of zeros of $f(\pi,p)+C_1(p)$ is the same as the set of Nash equilibrium profiles of the game $\Gamma$.
 In the following lemma we show that the set of zeros of $f(\pi,p)+C_1(p)$ is the same as the set of stationary points of the function $f(\pi,p)+C_1(p)$.\newline\newline
 \bf{Lemma 4.5: }\rm Given $(\pi^{*},p^{*})\in \Sigma\time\Sigma_{C}$. $f(\pi^{*},p^{*})+C_1(p^{*})=0$ iff $\nabla (f(\pi^{*},p^{*})+C_1(p^{*}))=0$.\newline
 \bf{Proof : }\rm [$\Rightarrow$] Since $f(\pi^{*},p^{*})+C_1(p^{*})=0$ and that $f$ and $C_1$ are non-negative, will imply that $f(\pi^{*},p^{*})=0$ and $C_1(p^{*})=0$.
 Thus, $\nabla f(\pi^{*},p^{*})=[\nabla_{\pi}f(\pi^{*},p^{*})^T\ \nabla_{p}f(\pi^{*},p^{*})^T]^T=0$ and $\nabla_{p}C_1(p^{*})=0$ by Lemma 4.3 and 4.4 respectively. Therefore, $\nabla (f(\pi^{*},p^{*})+C_1(p^{*})) = [\nabla_{\pi}f(\pi^{*},p^{*})^T\ (\nabla_{p}f(\pi^{*},p^{*})+\nabla_{p}C_1(p^{*}))^T]^T=0$.
 
 [$\Leftarrow$]Since $\nabla (f(\pi^{*},p^{*})+C_1(p^{*})) = [\nabla_{\pi}f(\pi^{*},p^{*})^T\ (\nabla_{p}f(\pi^{*},p^{*})+\nabla_{p}C_1(p^{*}))^T]^T=0$, we have $\nabla_{p}f(\pi^{*},p^{*})+\nabla_{p}C_1(p^{*})=0$.
 $(\nabla_{p}f(\pi^{*},p^{*})+\nabla_{p}C_1(p^{*}))^Tp^{*}=\nabla_{p}f(\pi^{*},p^{*})^Tp^{*}+\nabla_{p}C_1(p^{*})^Tp^{*}=0$. By substituting the expressions for $\nabla_{p}f(\pi^{*},p^{*})$ and $\nabla_{p}C_1(p^{*})$ we get,
 $\nabla_{p}f(\pi^{*},p^{*})^Tp^{*}=\{2\sum\limits_{i \in I}\sum\limits_{a\in A} (h^{i,a}(\pi^{*})^Tp^{*})h^{i,a}(\pi^{*})\}^Tp^{*}=2\sum\limits_{i \in I}\sum\limits_{a\in A} (h^{i,a}(\pi^{*})^Tp^{*})^2=2f(\pi^{*},p^{*})$ and $\nabla_{p}C_1(p^{*})^Tp^{*}=\{2\sum\limits_{i \in I}\sum\limits_{j=1}^{m_i}(max\{(g^{i,j})^Tp^{*},0\})g^{i,j}\}^Tp^{*}=2\sum\limits_{i \in I}\sum\limits_{j=1}^{m_i}(max\{(g^{i,j})^Tp^{*},0\})(g^{i,j})^Tp^{*}\\=2\sum\limits_{i \in I}\sum\limits_{j=1}^{m_i}(max\{(g^{i,j})^Tp^{*},0\})^2=2C_1(p^{*})$.
 Therefore , $0=(\nabla_{p}f(\pi^{*},p^{*})+\nabla_{p}C_1(p^{*}))^Tp^{*}=\nabla_{p}f(\pi^{*},p^{*})^Tp^{*}+\nabla_{p}C_1(p^{*})^Tp^{*}=2(f(\pi^{*},p^{*})+C_1(p^{*}))$.\hfill $\blacksquare$\newline
 
 Lemma 4.5 shows that the function $f(\pi,p)+C_1(p)$ is invex. Similarly it can shown that $f(\pi,p)+C_2(p)$ is also invex.
 
 In following lemma we show that $B$ is a biconvex function. As a consequence of this lemma, lemma 4.1 and lemma 3.3 in \cite{gorksi}, we get, $f(\pi,p)+B(\pi,p)$ is a biconvex function.\newline\newline
 \bf{Lemma 4.6: }\rm $B$ is a biconvex function i.e. $\forall \pi\in \Sigma,\ B(\pi,.):\Sigma_{C}\rightarrow[0,\infty)$ is a convex function and $\forall p\in \Sigma_{C},\ B(.,p):\Sigma\rightarrow[0,\infty)$ is a convex function.\newline
 \bf{Proof : }\rm Proof is similar to that of Lemma 4.1.\hfill $\blacksquare$\newline
 
 \section{Optimization problems.}
 In this section we shall state the optimization problems obtained using the functions constructed in the previous sections such that the global minima of the optimization problem correspond to Nash equilibria of the game $\Gamma$.
 
 First optimization problem ($O.P.1$) is stated below:
 
 \begin{align*}
  (O.P.1):\ \ \ \ \ \ \ \ \ \ \ &\min_{(\pi,p)}\ f(\pi,p)+C_1(p) \ \ \ \ \ \ \ \ \ \ \ \ \\
  & subject\ to:\ \ \ \ \ \ \ \ \ \ \ \ \ \ \ \ \ \ \ \ \ \ \ \ \ \ \ \\
  & \pi^i(a^i_{j})\geq0\ \ \ \ \ \ \ \ \ \ \ \forall i\in I,\ \forall j\in\{1,\dots,m_i\},\\
  & p(a)\geq0\ \ \ \ \ \ \ \ \ \ \ \ \ \forall a\in A,\\
  & \sum\limits_{j=1}^{m_i}\pi^i(a^i_j)=1\ \ \ \ \ \ \forall i\in I,\\
  & \sum\limits_{a\in A}p(a)=1.
 \end{align*}
 
 The constraints in the above optimization problem ensure that the feasible set is $\Sigma\times\Sigma_{C}$. The second optimization problem ($O.P.2$) is stated below: 
 
 \begin{align*}
  (O.P.2):\ \ \ \ \ \ \ \ \ \ \ &\min_{(\pi,p)}\ f(\pi,p)+B(\pi,p) \ \ \ \ \ \ \ \ \ \ \ \ \\
  & subject\ to:\ \ \ \ \ \ \ \ \ \ \ \ \ \ \ \ \ \ \ \ \ \ \ \ \ \ \ \\
  & \pi^i(a^i_{j})\geq0\ \ \ \ \ \ \ \ \ \ \ \forall i\in I,\ \forall j\in\{1,\dots,m_i\},\\
  & p(a)\geq0\ \ \ \ \ \ \ \ \ \ \ \ \ \forall a\in A,\\
  & \sum\limits_{j=1}^{m_i}\pi^i(a^i_j)=1\ \ \ \ \ \ \forall i\in I,\\
  & \sum\limits_{a\in A}p(a)=1.
 \end{align*}
 
The following theorem says that the set of global minima of the optimization problem ($O.P.1$) is the same as the set of Nash equilibria profiles of the game $\Gamma$.\newline\newline
\bf{Theorem 5.1: }\rm For every game $\Gamma$, there exists $(\pi^{*},p^{*})\in \Sigma\times\Sigma_{C}$ s.t. $f(\pi^{*},p^{*})+C_1(p^{*})=0$. Further given $(\pi^{*},p^{*})\in\Sigma\time\Sigma_{C}$, $f(\pi^{*},p^{*})+C_1(p^{*})=0$ iff $(\pi^{*},p^{*})$ is a Nash equilibrium profile.\newline
\bf{Proof : }\rm Since for every game there exists $\pi^{*}\in\Sigma$, s.t., $\pi^{*}$ is a N.E. (see \cite{nash}). Thus by theorem 3.2, $(\pi^{*},p^{*})$ with $p^{*}=P(\pi^{*})$ satisfies $f(\pi^{*},p^{*})+C_1(p^{*})=0$. The other part follows directly from theorem 3.2.\hfill $\blacksquare$\newline   

A similar claim can be proved for $O.P.2$.

The above two optimization problems have a biconvex objective function with convex (linear) constraints. Global optimization algorithm exists that solves the above two optimization problems (see \cite{floudas}). 

\section{The projected gradient descent algorithm and its convergence analysis.}

In this section we shall consider a projected gradient descent algorithm to solve $O.P.1$. The algorithm is stated below:\newline\newline 
\bf{Input:}\rm
 \begin{itemize}
  \item $<\pi_0,p_0>$ : initial point for the algorithm,
  \item $\Gamma$ : the underlying game,
  \item $\{a(n)\}_{n\geq1}$: step size sequences chosen as follows:
  \begin{itemize}
  \item $\forall n,\ a(n)>0$,
  \item $\sum_{\substack{n=1}}^{\infty}a(n)=\infty$,
  \item $\sum_{\substack{n=1}}^{\infty}a^2(n)<\infty$,
 \end{itemize}
  \item $H(\cdot)$ : projection operator ensuring that $(\pi,p)$ remains in $\Sigma\times\Sigma_{C}$.
  \end{itemize}
  \bf{Output : }\rm
  After sufficiently large number of iterations($lim$) the algorithm outputs the terminal strategy $(\pi^*,p^{*})$.\newline
\begin{align*}
  &\bf{The\ Algorithm\ :\ }\rm\\
  &n\leftarrow 0,\ the\ iteration\ index\ \ \ \ \ \ \ \ \ \ \ \ \ \ \ \ \ \ \ \ \ \ \ \ \ \ \ \ \ \ \ \ \ \ \ \ \ \ \ \ \ \ \ \ \ \ \ \ \ \ \  \ \ \ \ \ \ \ \ \ \ \ \ \ \\
  &\bf{while}\rm(n\leq lim)\newline\\
  &\ \ \ \begin{pmatrix}
          \pi_{n+1}\\
          p_{n+1}
         \end{pmatrix}
         =H\big{(}\begin{pmatrix}
          \pi_{n}\\
          p_{n}
         \end{pmatrix}
         -a(n)\begin{pmatrix}
          \nabla_{\pi} f(\pi_{n},p_{n})\\
          \nabla_p (f(\pi_{n},p_{n})+C_1(p_{n}))
         \end{pmatrix}\big{)}\\
  &\ \ \ n\leftarrow n+1\\
  &\bf{end\ while}\rm
\end{align*}

In what follows we shall present the convergence analysis of the above projected gradient descent algorithm. We shall analyse the behaviour of the above algorithm using the O.D.E. method presented in \cite{dupius}. In order to use the results from  \cite{dupius}, we need the gradient function to be lipschitz continuous on $\Sigma\times\Sigma_{C}$, which is proved in the following lemma.\newline\newline
\bf{Lemma 6.1: }\rm There exists $L>0$, s.t., $\forall\ (\pi_{1},p_{1}),\ (\pi_{2},p_{2})\in \Sigma\times\Sigma_{C}$,
\begin{center}
 $||\begin{pmatrix}
          \nabla_{\pi} f(\pi_{1},p_{1})\\
          \nabla_p (f(\pi_{1},p_{1})+C_1(p_{1}))
         \end{pmatrix}-\begin{pmatrix}
                       \nabla_{\pi} f(\pi_{2},p_{2})\\
                       \nabla_p (f(\pi_{2},p_{2})+C_1(p_{2}))
                       \end{pmatrix}||\leq L||\begin{pmatrix}
                                              \pi_{1}\\
                                              p_{1}
                                              \end{pmatrix}-\begin{pmatrix}
                                                            \pi_{2}\\
                                                            p_{2}
                                                            \end{pmatrix}||$
\end{center} 
\bf{Proof : } \rm It is easy to see that the function $f(\cdot)$ is twice continuously differentiable on an open set containing $\Sigma\times\Sigma_{C}$. Thus $\nabla f(\cdot)$ is continuously diffrentiable on $\Sigma\times\Sigma_{C}$. Hence $||\nabla^{2} f(\cdot)||\leq L_{1}$ for some $L_1>0$. By mean value theorem, we have, $\nabla f(\cdot)$ is Lipschitz continous with Lipschitz constant $L_1$. Let $\alpha:=\max\limits_{(i,j)\ :\ i\in I,\ a^i_j\in A^i}||g^{i,j}||$. 
Fix $(\pi_{1},p_{1}),\ (\pi_{2},p_{2})\in \Sigma\times\Sigma_{C}$. Clearly, $\forall i\in I,\ \forall j\in\{1,\dots,m_i\}, |max\{(g^{i,j})^Tp_1,0\}-max\{(g^{i,j})^Tp_2,0\}|\leq|(g^{i,j})^T(p_1-p_2)|$. Therefore, we have,
\begin{align*}
||\nabla C_1(p_1)-\nabla C_1(p_2)|| &\leq \sum\limits_{(i,j)}||g^{i,j}|||max\{(g^{i,j})^Tp_1,0\}-max\{(g^{i,j})^Tp_2,0\}|\\
&\leq \alpha\sum\limits_{(i,j)}|(g^{i,j})^T(p_1-p_2)|\\
&\leq \alpha\sum\limits_{(i,j)}||g^{i,j}||||(p_1-p_2)||\\
&\leq \alpha^2\sum\limits_{(i,j)}||(p_1-p_2)||\\
&= \alpha^2\beta||(p_1-p_2)||
\end{align*}
where $\beta=|\times_{i\in I}\big{[}\{i\}\times\{1,\dots,m_i\}\big{]}|$. Since $||p_1-p_2||=\sqrt{||p_1-p_2||^2}\leq \sqrt{||\pi_1-\pi_2||^2+||p_1-p_2||^2}$, we have, $||\nabla C_1(p_1)-\nabla C_1(p_2)||\leq L_2||(\pi_1,p_1)-(\pi_2,p_2)||$, where $L_2:=\alpha^2\beta$. 
Since sum of two lipschitz continuous functions is lipschitz continuous, we have, $\nabla(f(\cdot)+C_1(\cdot))$ is lipschitz continous with lipschitz constant $L:=L_1+L_2$.\hfill $\blacksquare$\newline  

In order to study the asymptotic behaviour of the recursion presented in the algorithm, by results in Section 3.4 of \cite{dupius}, it is enough to study the asymptotic behaviour of the o.d.e.,
\begin{equation}
\label{o.d.e.}
 \begin{pmatrix}
 \dot{\pi}\\
 \dot{p}
 \end{pmatrix}=\gamma\big{(}\begin{pmatrix}
                            \pi\\
                             p
                            \end{pmatrix};-\begin{pmatrix}
                                           \nabla_{\pi} f(\pi,p)\\
                                           \nabla_p (f(\pi,p)+C_1(p))
                                           \end{pmatrix}\big{)} 
\end{equation}
where $\forall v\in \Sigma\times\Sigma_{C},\ \forall d\in\mathbb{R}^{M_1+M_2},\ \gamma(v;d)=\lim\limits_{\delta\rightarrow 0}\frac{H(v+\delta d)-v}{\delta}$ i.e. the directional derivative of 
$H(\cdot)$ at $v$ along the direction $d$. The above o.d.e. is well posed i.e. has a unique solution for every initial point in $\Sigma\times\Sigma_{C}$ (for a proof see \cite{dupius}). 

$\Sigma\times\Sigma_{C}$, is a cartesian product of simplices and hence the projection of $(\hat{\pi},\hat{p})\in \mathbb{R}^{M_1+M_2}$ on to $\Sigma\times\Sigma_{C}$ is the same as projection of $\hat{\pi}^{i}$ on to $\Sigma^i,\ \forall i\in I$ and $\hat{p}$ on to $\Sigma_{C}$ i.e. $H((\hat{\pi}^T,\hat{p}^T)^T)=[H_{m_1}(\hat{\pi}^{1})^T,\dots,\ H_{m_N}(\hat{\pi}^{N})^T,\ H_{M_2}(\hat{p})^T]^T$ where $\forall n\in \mathbb{N},\ H_{n}(\cdot)$ denotes the projection operator which projects every vector in $\mathbb{R}^n$ on to $\bigtriangleup^{n}\subseteq\mathbb{R}^{n}$.
 Thus, in order to compute the directional derivative of $H(\cdot)$, it is enough to consider the directional derivative of the projection operator on to individual simplices and then juxtaposing them would give us the directional derivative of $H(\cdot)$.

The computation of the directional derivative of a projection operation on to a simplex can be found in \cite{dupius} which we shall state here. Let $\forall v\in \bigtriangleup^n,\ \forall d\in \mathcal{R}^n,\ \gamma_{n}(v;d):=\lim\limits_{\delta\rightarrow 0}\frac{H_n(v+\delta d)-v}{\delta}$ and $\eta(v):=\{x\in \mathbb{R}^n\ :\ ||x||=1,\ \langle x,v-\hat{v}\rangle\leq0,\ \forall \hat{v}\in \bigtriangleup^n\}$. Then, 
\begin{equation}
\label{dd}
 \gamma_{n}(v;d)=d+(max\{\langle d,-x_n\rangle,0\})x_n
\end{equation}
where $x_n\in \eta(v)$, s.t., $\forall x\in\eta(v),\ \langle d,-x_n\rangle\ \geq\ \langle d,-x\rangle$. 

Let $\forall (\pi,p)\in \Sigma\times\Sigma_{C},\ V(\pi,p):=f(\pi,p)+C_1(p)$. Fix $(\pi_0,p_0) \in \Sigma\times\Sigma_{C}$ be a initial point of the o.d.e. \ref{o.d.e.} and the corresponding unique solution be $(\pi(t),p(t))$. Then, 
\begin{align*}
\frac{dV(\pi(t),p(t))}{dt}&=\nabla V(\pi(t),p(t))^T\gamma\big{(}\small{\begin{pmatrix}
                            \pi\\
                             p
                            \end{pmatrix};-\begin{pmatrix}
                                           \nabla_{\pi} f(\pi,p)\\
                                           \nabla_p (f(\pi,p)+C_1(p))
                                           \end{pmatrix}}\big{)}\\
                         &=\sum\limits_{i\in I}\nabla_{\pi^i}V(\pi(t),p(t))^T\gamma_{m_i}\big{(}
                            \pi^i;-\nabla_{\pi^i} (f(\pi,p)+C_1(p))\big{)}\\
                         &+\nabla_{p}V(\pi(t),p(t))^T\gamma_{M_2}\big{(}
                            p;-\nabla_{p} (f(\pi,p)+C_1(p))\big{)}\\
\end{align*}

By substituing \ref{dd} and the fact that $\forall (\pi,p)\in \Sigma\times\Sigma_{C},\ \nabla V(\pi,p)=\nabla(f(\pi,p)+C_1(p))$ in the above equation we get,
\begin{align*}
\frac{dV(\pi(t),p(t))}{dt}\leq&\sum\limits_{i\in I}(-||\nabla_{\pi^i} f(\pi,p)||^2 + |\langle\nabla_{\pi^i} f(\pi,p),x_{m_i}\rangle|^2)\\
                              &+(-||\nabla_{p}(f(\pi,p)+C_1(p))||^2 + |\langle\nabla_{p}(f(\pi,p)+C_1(p)),x_{M_2}\rangle|^2)\\
                           \leq&0.   
\end{align*}
where the last inequality follows from the application of cauchy schwartz and the fact that $\forall n\in \mathbb{N},\ ||x_n||=1$.

Therefore along every solution of the o.d.e. \ref{o.d.e.}, the value of the potential function $V(\cdot)$ reduces and hence the above o.d.e. converges to an internally chain transitive invariant set contained in $\mathcal{L}:=\{(\pi^*,p^*)\in \Sigma\times\Sigma_{C}\ :\ \frac{dV(\pi^*,p^*)}{dt}=0\}$.

In the following lemma we shall prove that $(\pi^*,p^*)\in \mathcal{L}$ is an equilibrium point of o.d.e. \ref{o.d.e.}.\newline\newline
\bf{Lemma 6.2: }\rm If $(\pi^*,p^*)\in \mathcal{L}$, then, $\gamma\big{(}\small{\begin{pmatrix}\pi^*\\p^*\end{pmatrix};-\begin{pmatrix}\nabla_{\pi} f(\pi^*,p^*)\\\nabla_p (f(\pi^*,p^*)+C_1(p^*))\end{pmatrix}}\big{)}=0$.\newline
\bf{Proof : }\rm If $(\pi^*,p^*)\in \mathcal{L}$ is such that $\nabla(f(\pi^*,p^*)+C_1(p^*))=0$, then $\gamma((\pi^*,p^*);\nabla(f(\pi^*,p^*)+C_1(p^*)))=0$. Assume $\nabla(f(\pi^*,p^*)+C_1(p^*))\neq0$. Since $(\pi^*,p^*)\in \mathcal{L}$, $\sum\limits_{i\in I}(-||\nabla_{\pi^i} f(\pi,p)||^2 + |\langle\nabla_{\pi^i} f(\pi,p),x_{m_i}\rangle|^2)+(-||\nabla_{p}(f(\pi,p)+C_1(p))||^2 + |\langle\nabla_{p}(f(\pi,p)+C_1(p)),x_{M_2}\rangle|^2)=0$. By cauchy schwartz inequality, $\forall i\in I,\ (-||\nabla_{\pi^i} f(\pi,p)||^2 + |\langle\nabla_{\pi^i} f(\pi,p),x_{m_i}\rangle|^2)\leq0$ and $(-||\nabla_{p}(f(\pi,p)+C_1(p))||^2 + |\langle\nabla_{p}(f(\pi,p)+C_1(p)),x_{M_2}\rangle|^2)\leq0$. Since their sum is zero, we get, $\forall i\in I,\ (-||\nabla_{\pi^i} f(\pi,p)||^2 + |\langle\nabla_{\pi^i} f(\pi,p),x_{m_i}\rangle|^2)=0$ and $(-||\nabla_{p}(f(\pi,p)+C_1(p))||^2 + |\langle\nabla_{p}(f(\pi,p)+C_1(p)),x_{M_2}\rangle|^2)=0$. Hence, $\forall i\in I,\ x_{m_i}=\pm \frac{\nabla_{\pi^i}f(\pi^*,p^*)}{||\nabla_{\pi^i}f(\pi^*,
p^*)||}$ and $x_{M_2}=\pm \frac{\nabla_{p}(f(\pi^*,p^*)+C_1(p^*))}{||\nabla_{p}(f(\pi^*,p^*)+C_1(p^*))||}$. 
By, definition of $x_n$ in equation \ref{dd}, we get, $\forall i\in I,\ x_{m_i}=\frac{\nabla_{\pi^i}f(\pi^*,p^*)}{||\nabla_{\pi^i}f(\pi^*,p^*)||}$ and $x_{M_2}=\frac{\nabla_{p}(f(\pi^*,p^*)+C_1(p^*))}{||\nabla_{p}(f(\pi^*,p^*)+C_1(p^*))||}$. Substituing for $x_{m_i}$ and $x_{M_2}$ in the expression for $\gamma_{m_i}((\pi^i)^*;-\nabla_{\pi^i}(f(\pi^*,p^*)+C_1(p^*)))$ and $\gamma_{M_2}(p^*;-\nabla_{p}(f(\pi^*,p^*)+C_1(p^*)))$ and using the fact that $\gamma((\pi^*,p^*);-\nabla(f(\pi^*,p^*)+C_1(p^*)))=[(\gamma_{m_1}((\pi^1)^*;-\nabla_{\pi^1}(f(\pi^*,p^*)+C_1(p^*))))^T,\dots,\ (\gamma_{m_N}((\pi^N)^*;-\nabla_{\pi^N}(f(\pi^*,p^*)+C_1(p^*))))^T,\ (\gamma_{M_2}(p^*;-\nabla_{p}(f(\pi^*,p^*)+C_1(p^*))))^T]^T$ we get the desired result.\hfill $\blacksquare$\newline

In fact the converse is also true and the proof is similar to that of the previous lemma. Therefore $\mathcal{L}=\mathcal{E}$ where $\mathcal{E}$ denotes the set of equilibrium points of o.d.e.\ref{o.d.e.}.
The following lemma says that every point in the set $\mathcal{L}$ is a partial optimum of the biconvex function $f(\pi,p)+C_1(p)$.\newline\newline
\bf{Lemma 6.3: }\rm $(\pi^*,p^*) \in \mathcal{L}$, then, $\forall \pi \in \Sigma,\ f(\pi^*,p^*)+C_1(p^*)\leq f(\pi,p^*)+C_1(p^*)$ and $\forall p \in \Sigma_{C},\ f(\pi^*,p^*)+C_1(p^*)\leq f(\pi^*,p)+C_1(p)$.\newline
\bf{Proof : }\rm If $(\pi^*,p^*) \in \mathcal{L}$ is such that $\nabla(f(\pi^*,p^*)+C_1(p^*))=0$, then by lemma 4.5 the result follows. Assume $\nabla(f(\pi^*,p^*)+C_1(p^*))\neq0$. Then by lemma 6.2 we have, $\forall i\in I,\ x_{m_i}=\frac{\nabla_{\pi^i}f(\pi^*,p^*)}{||\nabla_{\pi^i}f(\pi^*,p^*)||}$ and $x_{M_2}=\frac{\nabla_{p}(f(\pi^*,p^*)+C_1(p^*))}{||\nabla_{p}(f(\pi^*,p^*)+C_1(p^*))||}$.

By equation \ref{dd}, $x_{M_2}\in \eta(p^*)$ and hence $\forall p\in\Sigma_{C},\ \langle \frac{\nabla_{p}(f(\pi^*,p^*)+C_1(p^*))}{||\nabla_{p}(f(\pi^*,p^*)+C_1(p^*))||},p^*-p\rangle\leq0$. Therefore $\forall p\in \Sigma_{C},\ \langle \nabla_{p}(f(\pi^*,p^*)+C_1(p^*)),p - p^*\rangle\geq0$. By convexity of $f(\pi^*,\cdot)+C_1(\cdot)$ and proposition 1.1.8 in \cite{bert}, we get $\forall p \in \Sigma_{C},\ f(\pi^*,p^*)+C_1(p^*)\leq f(\pi^*,p)+C_1(p)$.

By equation \ref{dd}, $\forall i\in I,\ x_{m_i}\in \eta(\pi^{i*})$ and hence $\forall i\in I,\ \forall \pi^i\in \Sigma^i,\ \langle \frac{\nabla_{\pi^i}f(\pi^*,p^*)}{||\nabla_{\pi^i}f(\pi^*,p^*)||},(\pi^i)^*-\pi^i\rangle\leq0$. Therefore $\forall i\in I,\ \forall \pi^i\in \Sigma^i,\ \langle \nabla_{\pi^i}f(\pi^*,p^*),\pi^i-(\pi^i)^*\rangle\geq0$. Since $\forall \pi\in \Sigma,\ \langle \nabla_{\pi}(f(\pi^*,p^*)+C_1(p^*)),\pi-\pi^*\rangle=\langle \nabla_{\pi}f(\pi^*,p^*),\pi-\pi^*\rangle=\sum\limits_{i\in I}\langle \nabla_{\pi^i}f(\pi^*,p^*),\pi^i-(\pi^i)^*\rangle$, we get, $\forall \pi\in \Sigma,\ \langle \nabla_{\pi}(f(\pi^*,p^*)+C_1(p^*)),\pi-\pi^*\rangle\geq0$. Thus by convexity of $f(\cdot,p^*)+C_1(p^*)$ and by proposition 1.1.8 in \cite{bert}, we have, $\forall \pi\in \Sigma,\ f(\pi^*,p^*)+C_1(p^*)\leq f(\pi,p^*)+C_1(p^*)$.\hfill $\blacksquare$\newline  

Even though the proof guarantees convergence to the set of partial optimum of the biconvex function in simulation on various test cases it was observed that the iterates converge to the set of Nash equilibria of the game $\Gamma$.

\section{Simulation results.}

In the simulations carried out, in order to perform the projection operation in every iteration we use the procedure in \cite{proj}.

\subsection{Rock-Paper-Scissor : }
We consider the following version of the standard rock-paper-scissor game.
\setlength{\arraycolsep}{0pt}
\[
\begin{array}{cccccccccc}
\cline{2-9} \vline{} & \quad     & \vline{} & \quad R      \quad & \vline{} & \quad P      \quad & \vline{} & \quad S      \quad & \vline{}\\
\cline{2-9} \vline{} & \quad R \ & \vline{} & \quad (0,0)  \quad & \vline{} & \quad (0,1)  \quad & \vline{} & \quad (1,0)  \quad & \vline{}\\
\cline{2-9} \vline{} & \quad P \ & \vline{} & \quad (1,0)  \quad & \vline{} & \quad (0,0)  \quad & \vline{} & \quad (0,1)  \quad & \vline{}\\
\cline{2-9} \vline{} & \quad S \ & \vline{} & \quad (0,1)  \quad & \vline{} & \quad (1,0)  \quad & \vline{} & \quad (0,0)  \quad & \vline{}\\
\cline{2-9}
\end{array}
\]
In the above game, $((\frac{1}{3},\ \frac{1}{3},\ \frac{1}{3}),(\frac{1}{3},\ \frac{1}{3},\ \frac{1}{3}))$ is the only Nash equilibrium strategy. Having started the algorithm from a random initial point, variation of the objective function value and the strategies are shown in the plots below.
\begin{figure}[h]
 \begin{center}
 
 \begin{subfigure}[b]{0.47\textwidth}
  \includegraphics[width=\textwidth]{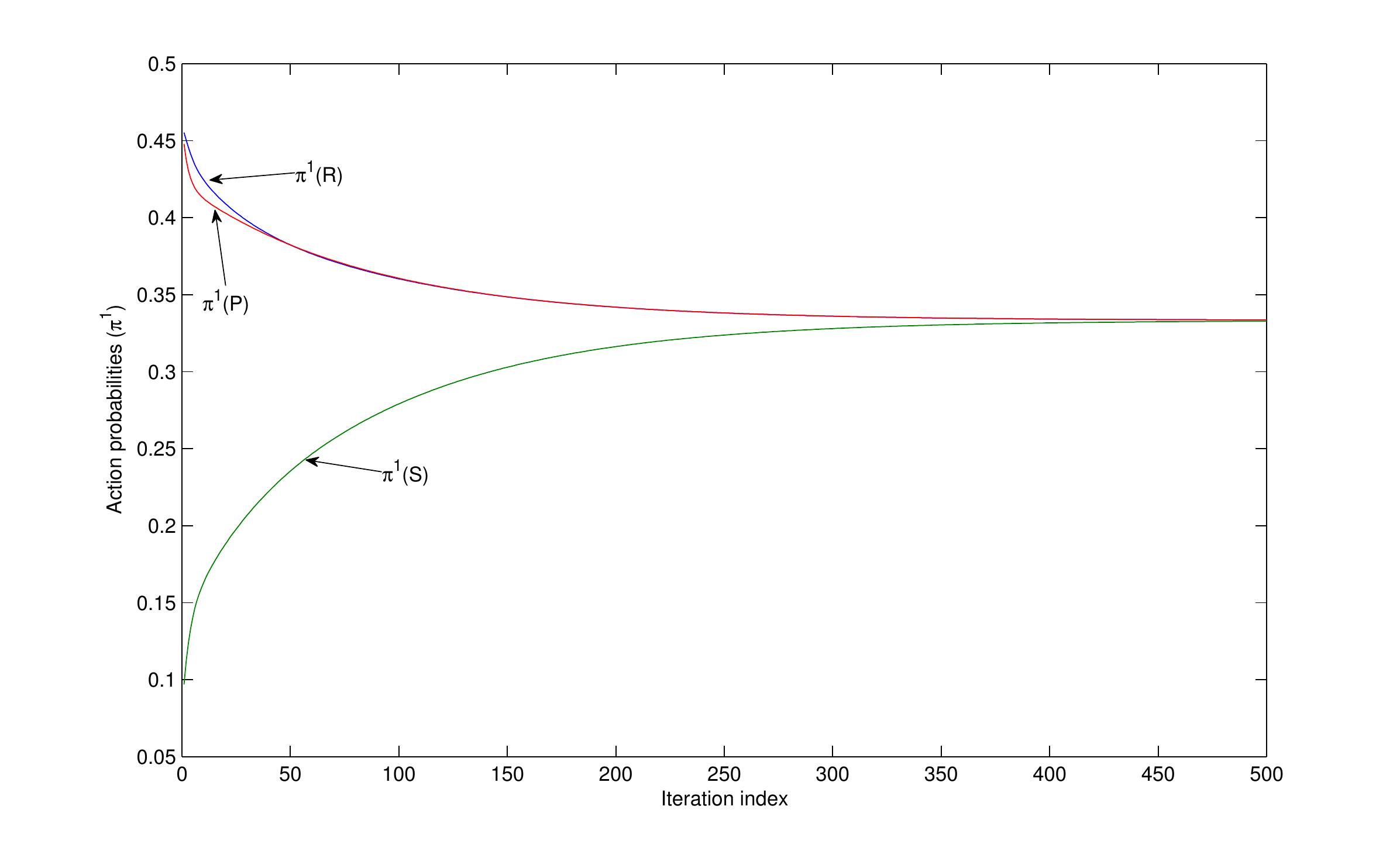}
  \caption{Action probabilities of player 1 vs iteration index.}
  \label{fig:RPS_p1}
 \end{subfigure}
 \begin{subfigure}[b]{0.495\textwidth}
  \includegraphics[width=\textwidth]{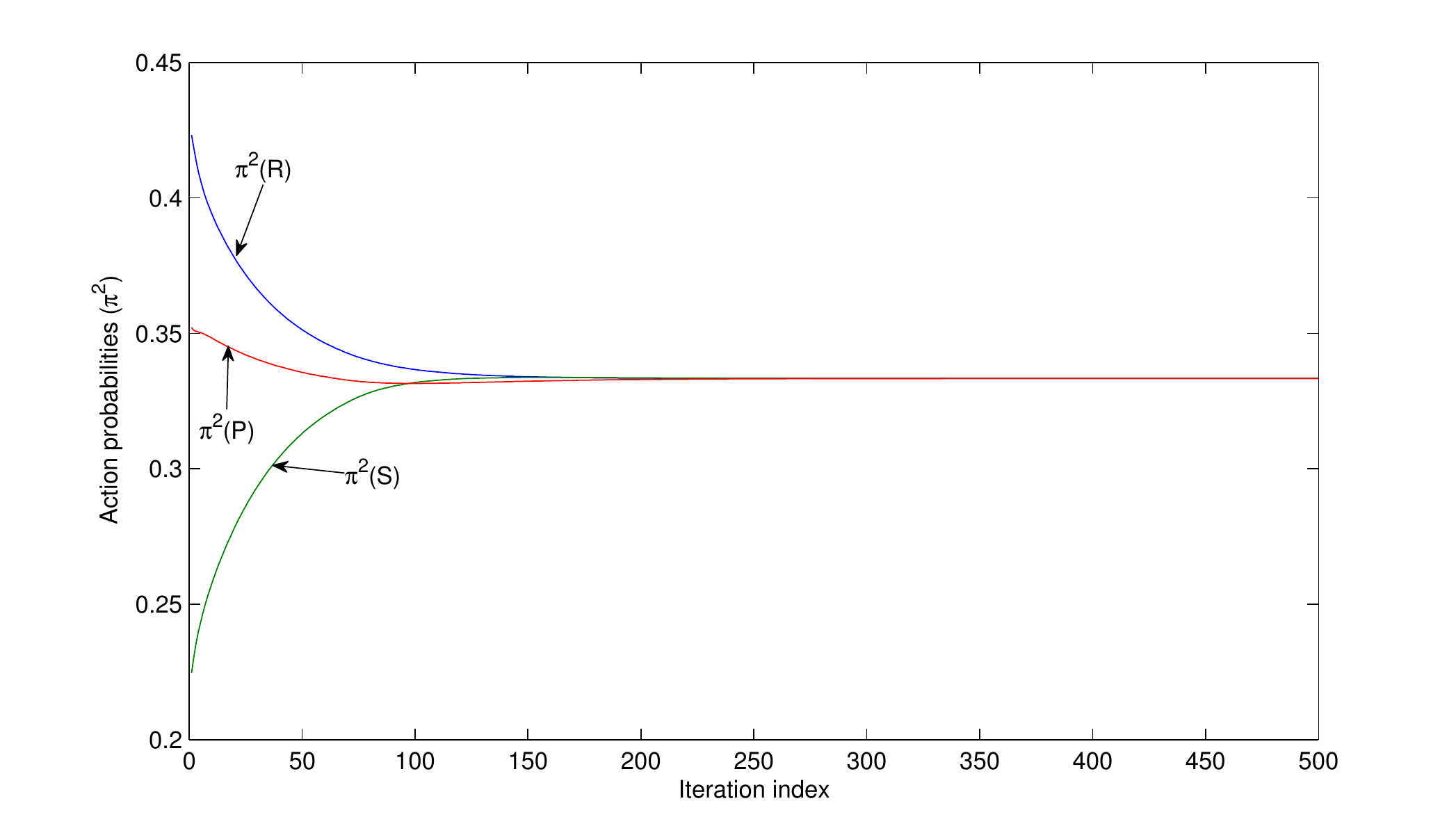}
  \caption{Action probabilities of player 2 vs iteration index.}
  \label{fig:RPS_p1}
 \end{subfigure}
\caption{Action probabilities vs iteration index}
\label{fig:RPS_ap}
\end{center}
\end{figure} 

\begin{figure}[h]
\begin{center}
\scalebox{0.5}{\includegraphics{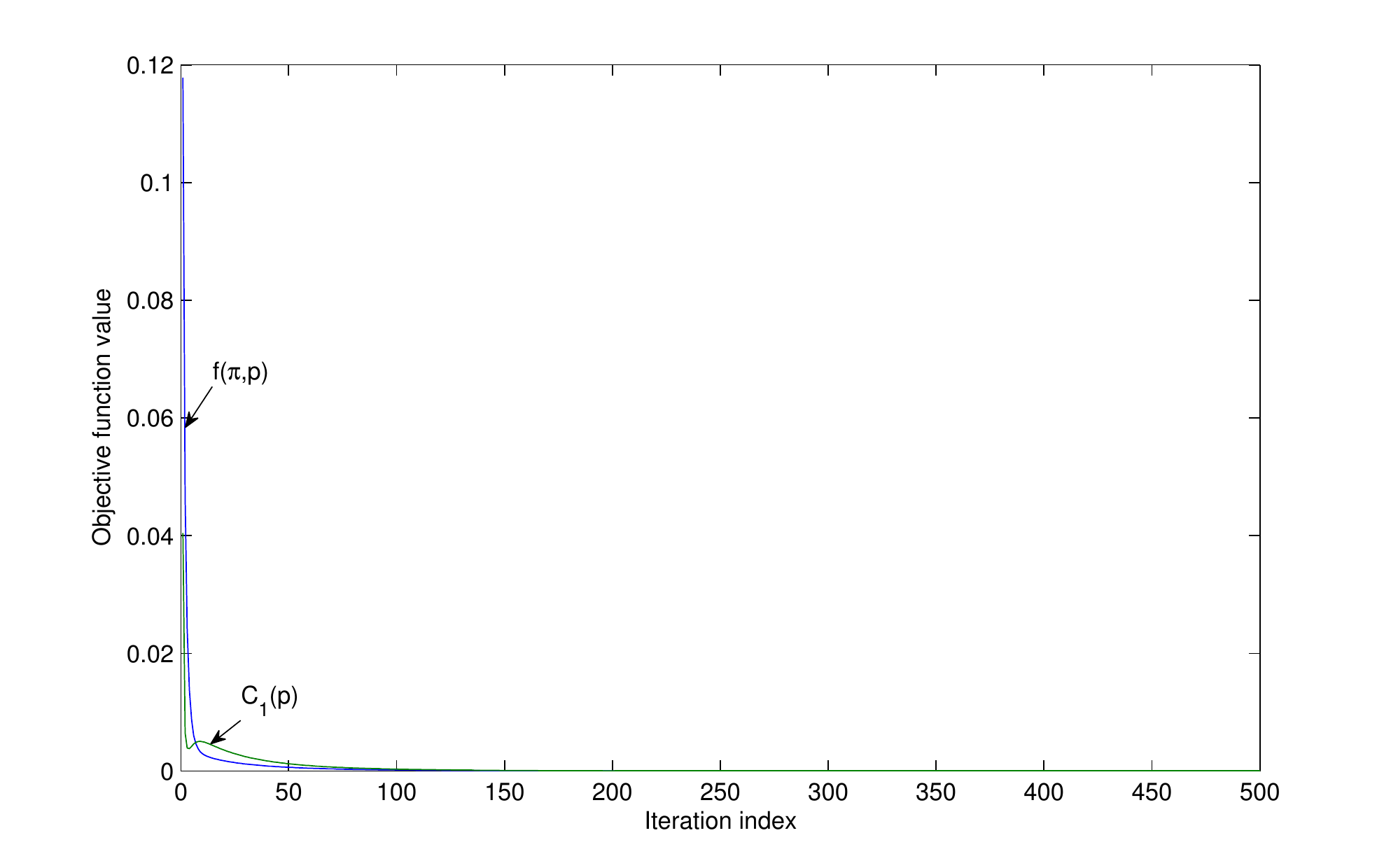}}
\caption{Objective function value vs iteration index.}
\label{fig:RPS_obj}
\end{center}
\end{figure}

The plots in Fig:\ref{fig:RPS_ap} show that the action probabilities converge to the Nash equilibrium of the game. As the action probabilities converge to Nash equilibrium strategy the objective function value approaches zero as seen in Fig:\ref{fig:RPS_obj}. 

\newpage
\subsection{Jordan's game : }
\begin{figure}[h]
 \begin{center}
 
 \begin{subfigure}[b]{0.435\textwidth}
  \includegraphics[width=\textwidth]{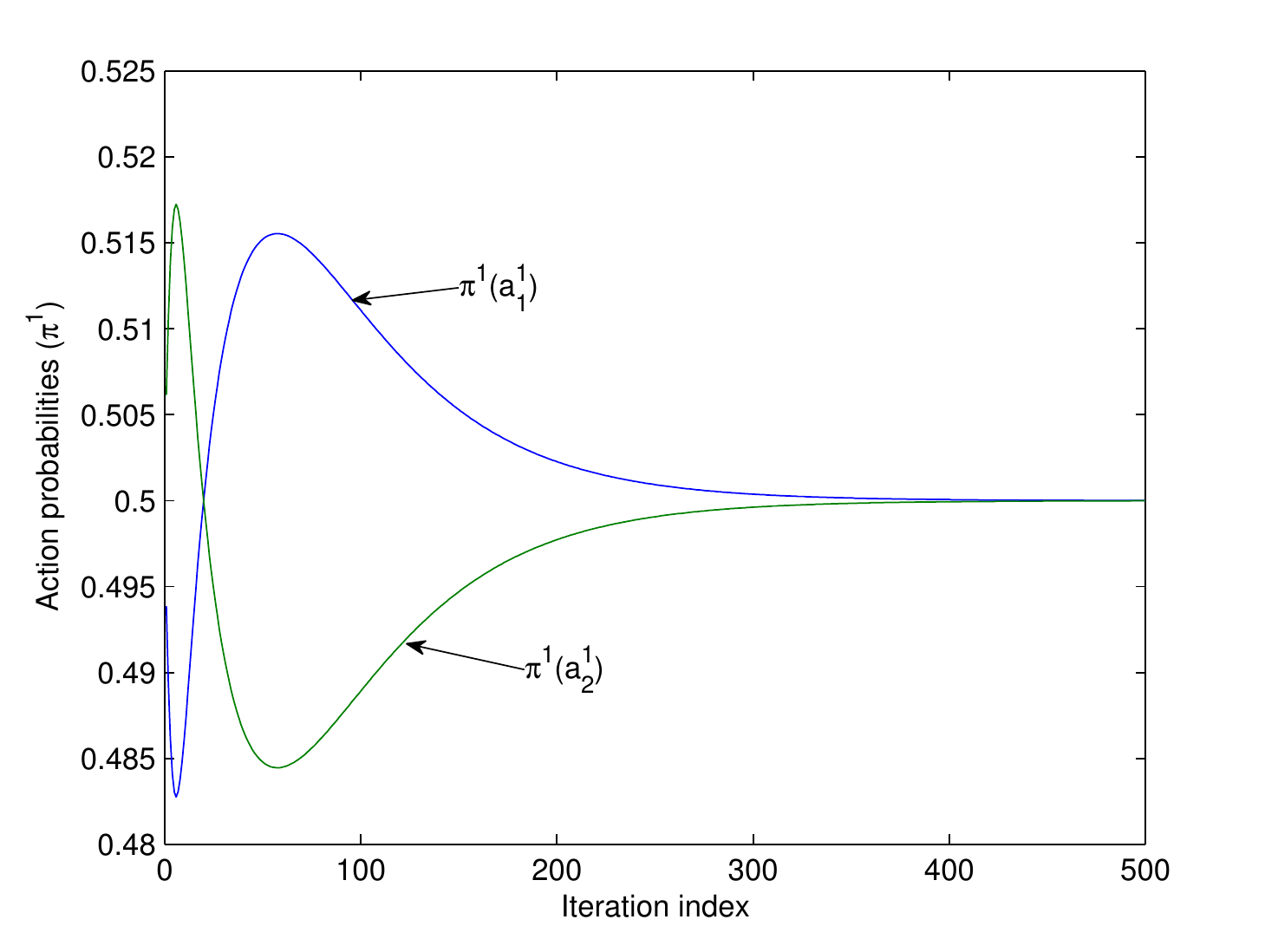}
  \caption{Action probabilities of player 1 vs iteration index.}
  \label{fig:jg_p1}
 \end{subfigure}
 \begin{subfigure}[b]{0.485\textwidth}
  \includegraphics[width=\textwidth]{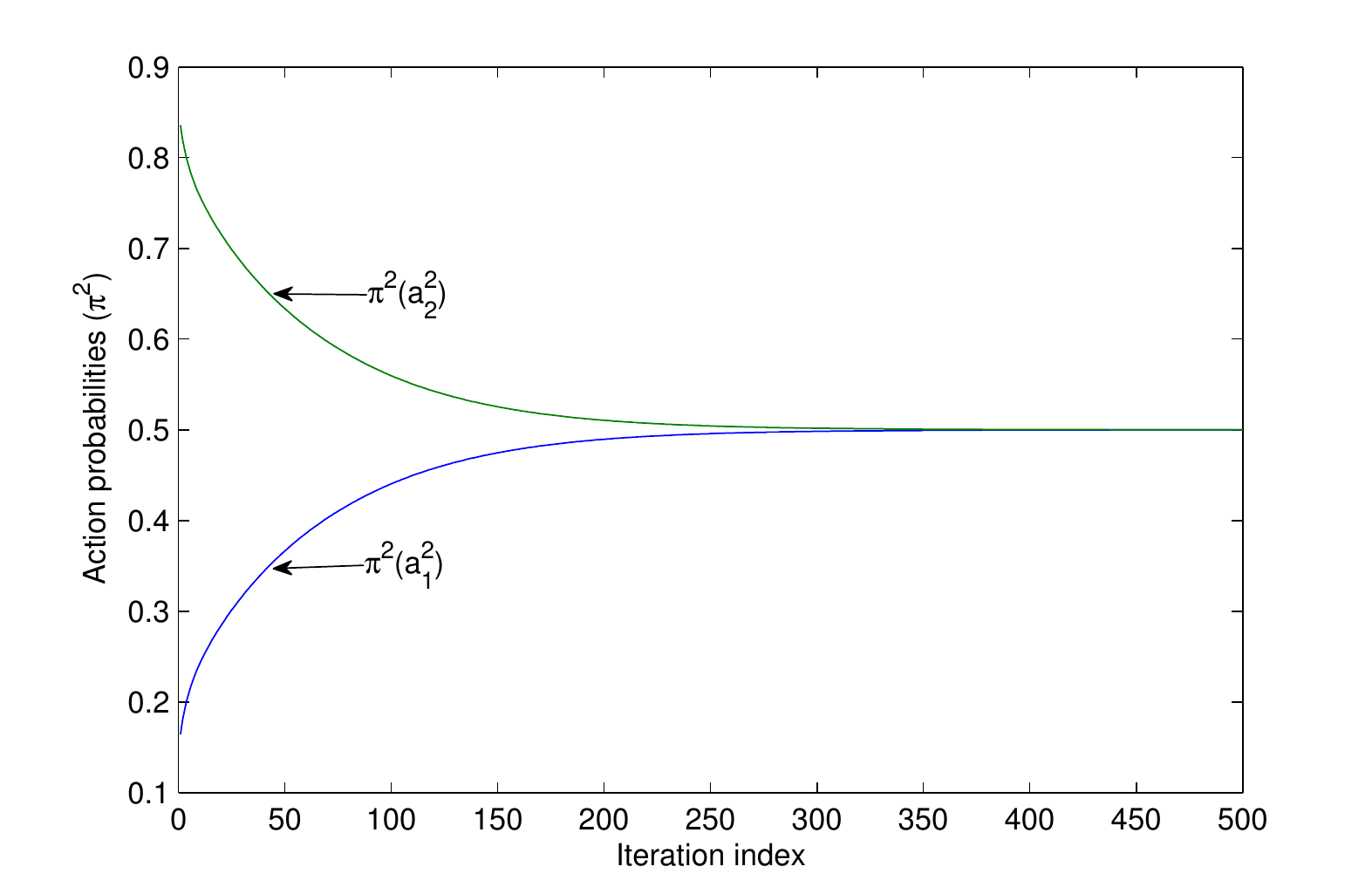}
  \caption{Action probabilities of player 2 vs iteration index.}
  \label{fig:jg_p2}
 \end{subfigure}
\caption{Action probabilities vs iteration index}
\label{fig:jg_ap}
\end{center}
\end{figure}

\begin{figure}[h]
 \begin{center}
 
 \begin{subfigure}[b]{0.435\textwidth}
  \includegraphics[width=\textwidth]{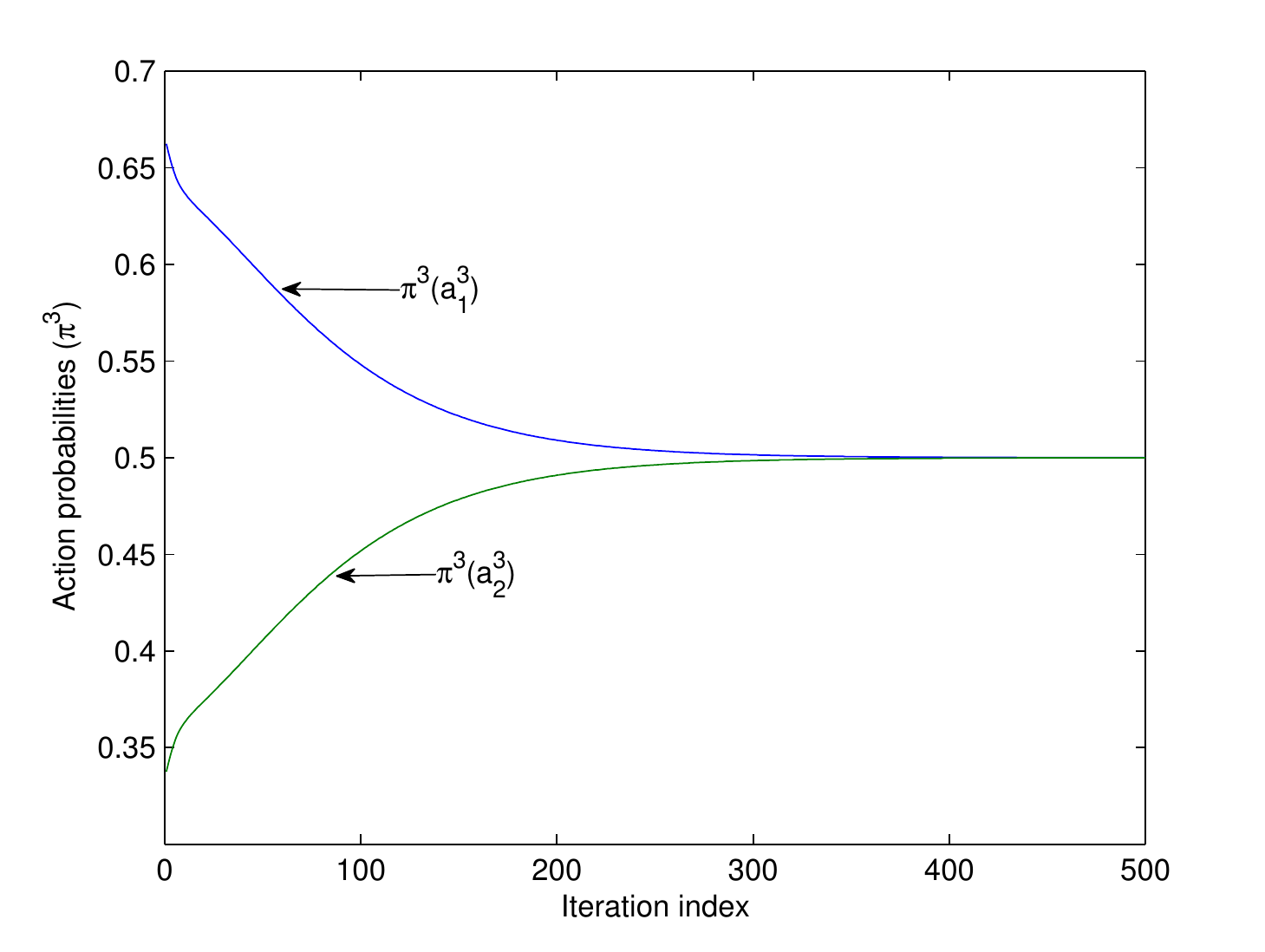}
  \caption{Action probabilities of player 3 vs iteration index.}
  \label{fig:jg_p3}
 \end{subfigure}
 \begin{subfigure}[b]{0.45\textwidth}
  \includegraphics[width=\textwidth]{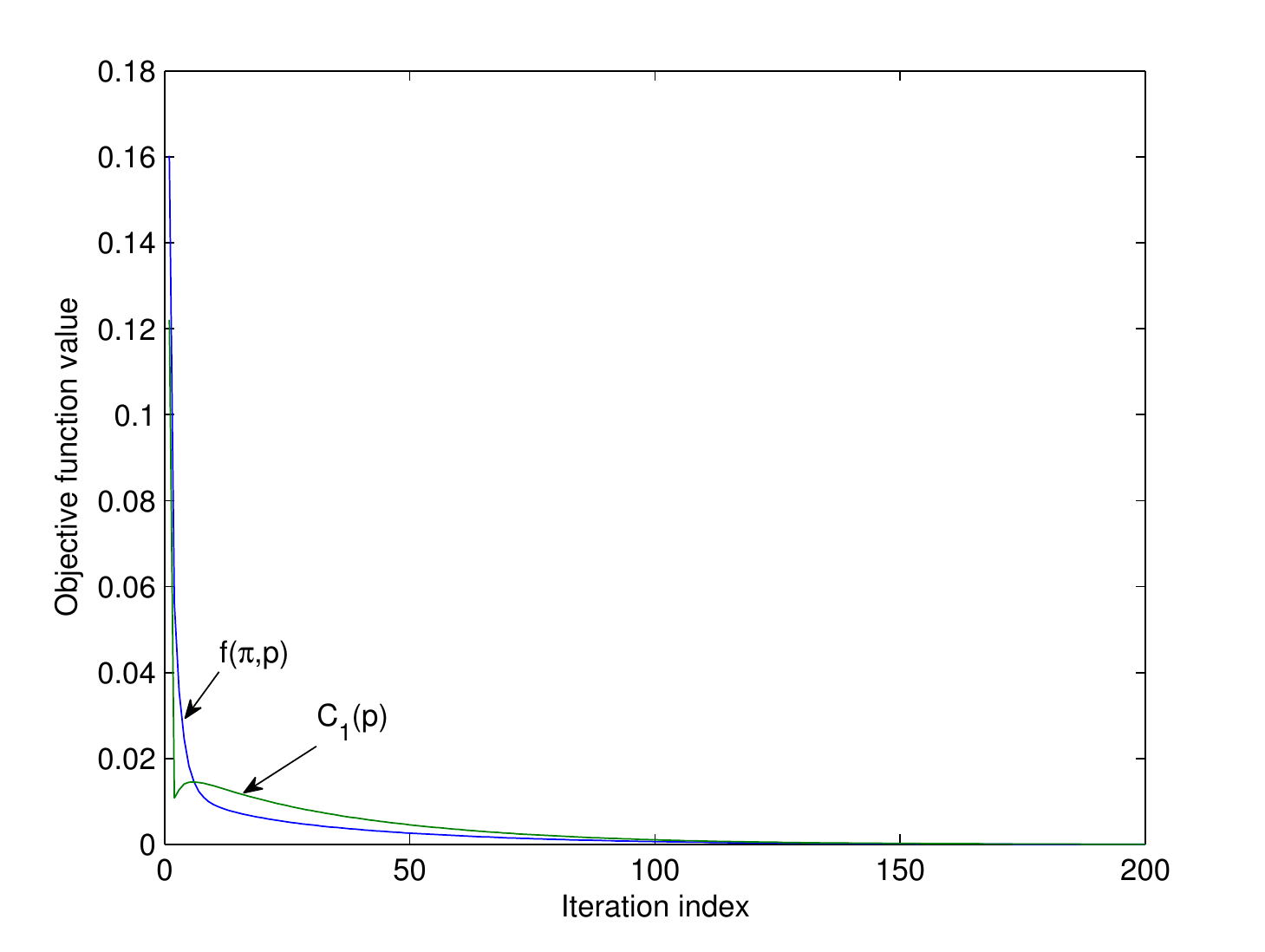}
  \caption{Objective function value vs iteration index.}
  \label{fig:jg_obj}
 \end{subfigure}
\caption{Action probabilities/Objective function value  vs iteration index}
\label{fig:jg_ap_obj}
\end{center}
\end{figure}

The general form of Jordan's game can be found in \cite{sergiu}. We consider the following version.
\begin{itemize}
 \item [] Player 3 action $a^3_1$:
 \setlength{\arraycolsep}{0pt}
\[
\begin{array}{ccccccc}
\cline{2-6} \vline{} & \quad     & \vline{} & \quad a^2_1      \quad & \vline{} & \quad a^2_2      \quad & \vline{} \\
\cline{2-6} \vline{} & \quad a^1_1 \ & \vline{} & \quad (0,0,0)  \quad & \vline{} & \quad (1,1,0) \quad & \vline{} \\
\cline{2-6} \vline{} & \quad a^1_2 \ & \vline{} & \quad (1,0,1) \quad & \vline{} & \quad (0,1,1)  \quad & \vline{} \\
\cline{2-6}
\end{array}
\]
\item [] Player 3 action $a^3_2$ :
\[
\begin{array}{ccccccc}
\cline{2-6} \vline{} & \quad     & \vline{} & \quad a^2_1      \quad & \vline{} & \quad a^2_2      \quad & \vline{} \\
\cline{2-6} \vline{} & \quad a^1_1 \ & \vline{} & \quad (0,1,1)  \quad & \vline{} & \quad (1,0,1) \quad & \vline{} \\
\cline{2-6} \vline{} & \quad a^1_2 \ & \vline{} & \quad (1,1,0) \quad & \vline{} & \quad (0,0,0)  \quad & \vline{} \\
\cline{2-6}
\end{array}
\]
\end{itemize}
In the above game, $((\frac{1}{2},\ \frac{1}{2}),(\frac{1}{2},\ \frac{1}{2}),(\frac{1}{2},\ \frac{1}{2}))$ is the only Nash equilibrium strategy. Having started the algorithm from a random initial point, variation of the objective function value and the strategies are shown in the plots in Fig:\ref{fig:jg_ap} and Fig:\ref{fig:jg_ap_obj}.

Simulations were also carried out on other versions of this game obtained from the general form in \cite{sergiu} and convergence to Nash equilibrium was observed.

\subsection{A game with finite number of Nash equilibria : }
\begin{figure}[h]
 \begin{center}
 
 \begin{subfigure}[b]{0.428\textwidth}
  \includegraphics[width=\textwidth]{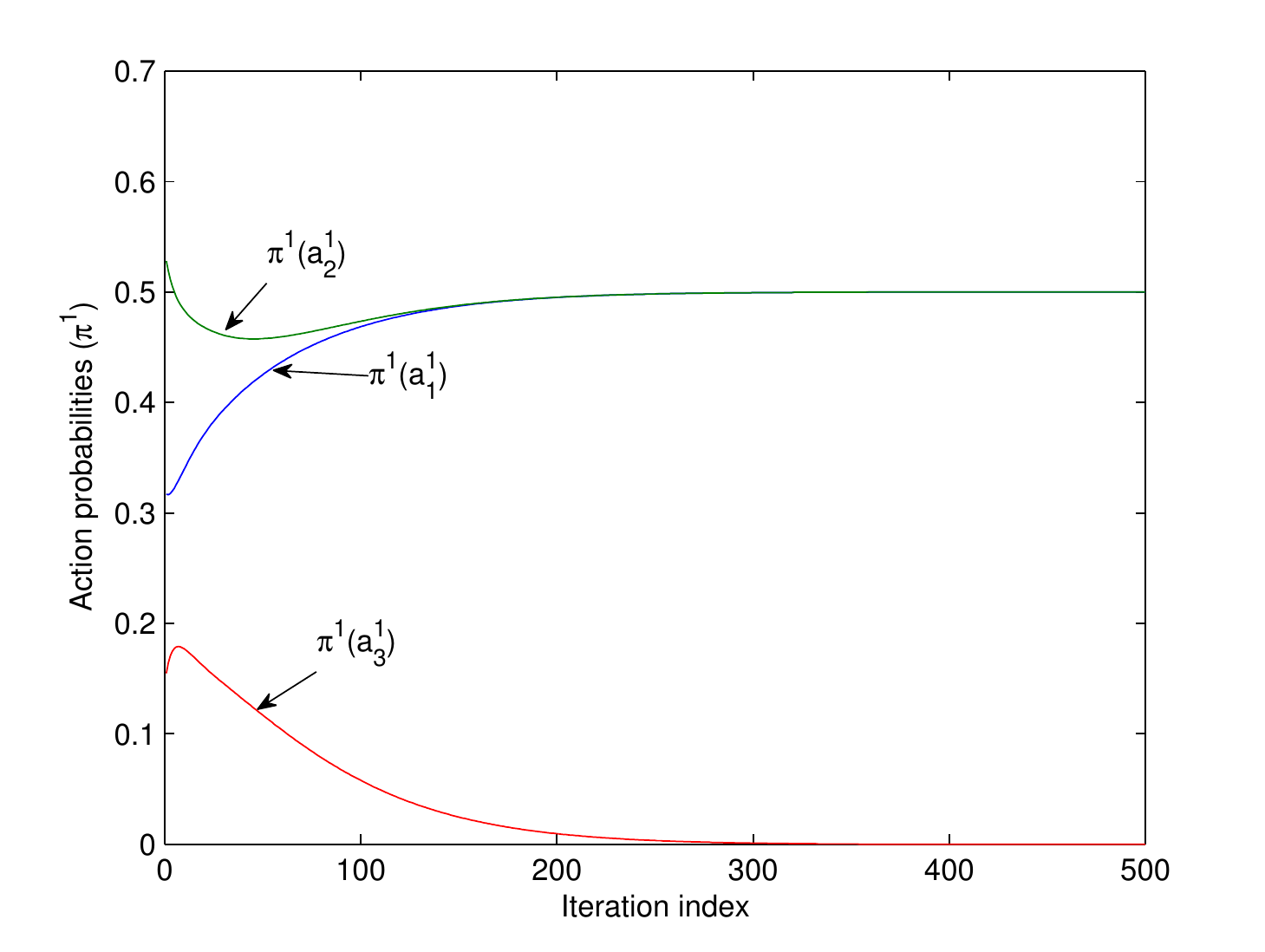}
  \caption{Action probabilities of player 1 vs iteration index.}
  \label{fig:hm_p1}
 \end{subfigure}
 \begin{subfigure}[b]{0.428\textwidth}
  \includegraphics[width=\textwidth]{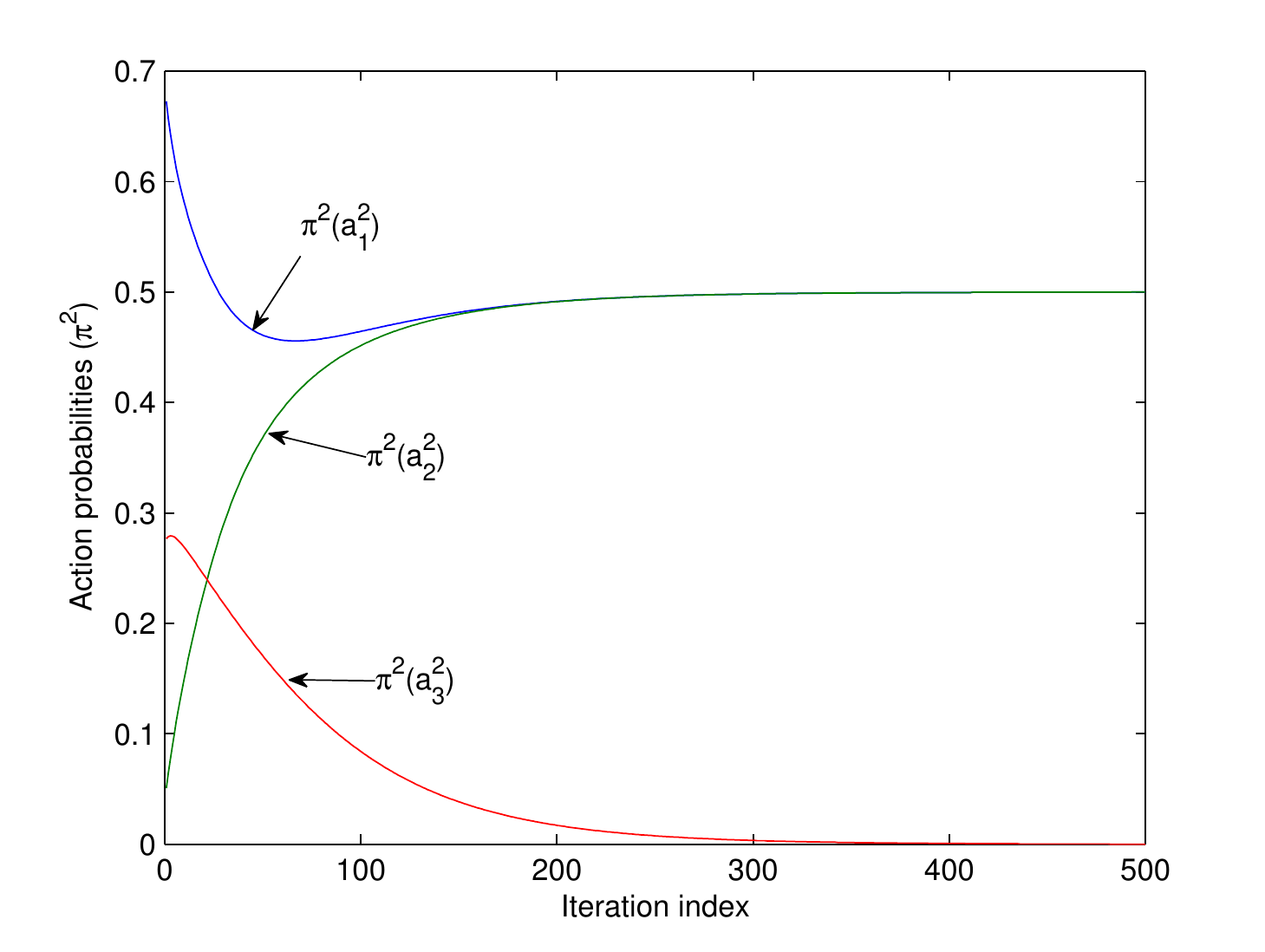}
  \caption{Action probabilities of player 2 vs iteration index.}
  \label{fig:hm_p2}
 \end{subfigure}
\caption{Action probabilities vs iteration index}
\label{fig:hm_ap}
\end{center}
\end{figure}

\begin{figure}[h]
\begin{center}
\scalebox{0.5}{\includegraphics{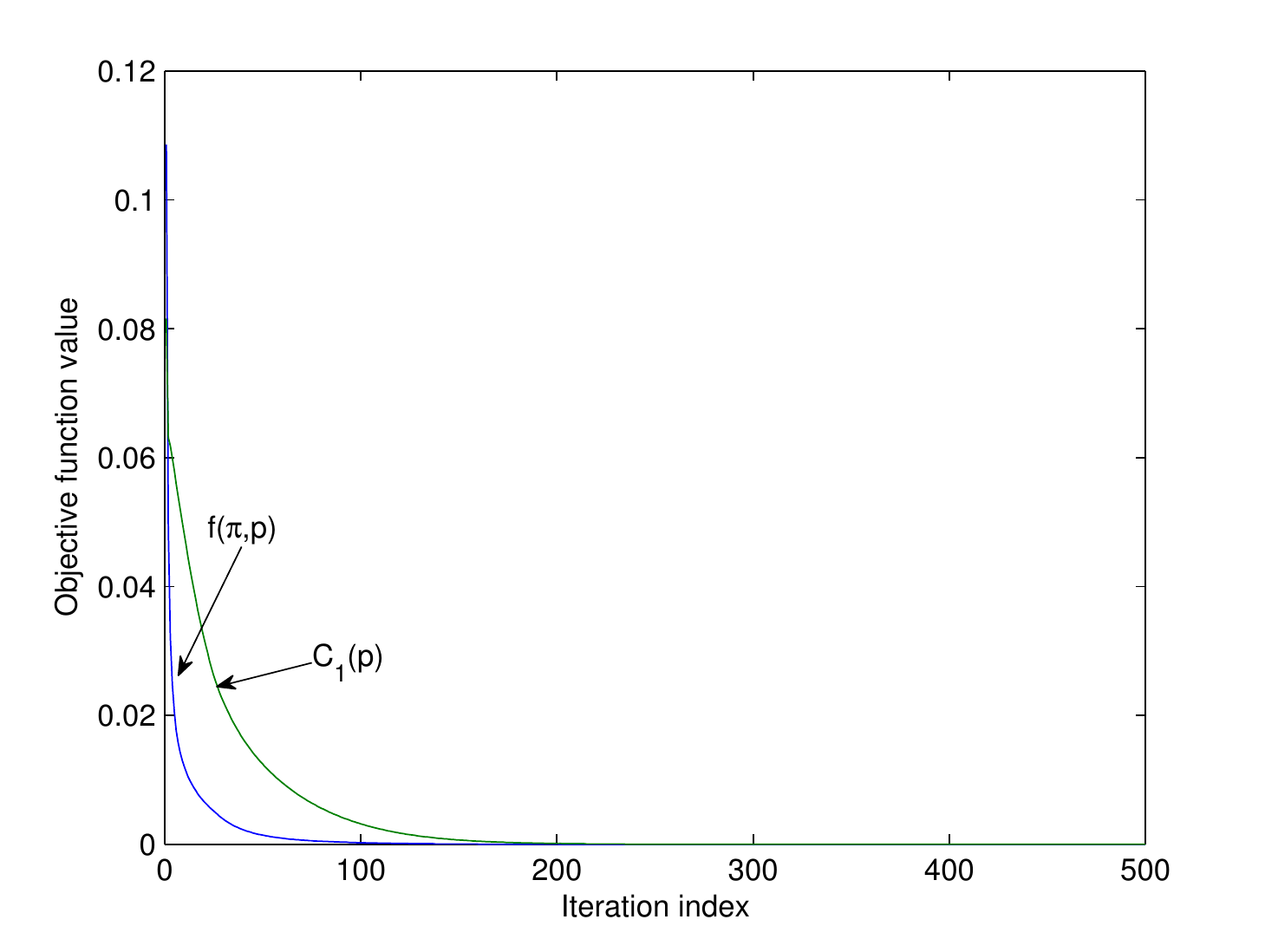}}
\caption{Objective function value vs iteration index.}
\label{fig:hm_obj}
\end{center}
\end{figure}

The following game was introduced in \cite{sergiu2} in order to show non-convergence of certain class of algorithms. The game is stated below.
 \setlength{\arraycolsep}{0pt}
\[
\begin{array}{cccccccccc}
\cline{2-9} \vline{} & \quad         & \vline{} & \quad a^2_1    \quad & \vline{} & \quad a^2_2    \quad & \vline{} & \quad a^2_3    \quad & \vline{}\\
\cline{2-9} \vline{} & \quad a^1_1 \ & \vline{} & \quad (1,0)    \quad & \vline{} & \quad (0,1)    \quad & \vline{} & \quad (1,0)    \quad & \vline{}\\
\cline{2-9} \vline{} & \quad a^1_2 \ & \vline{} & \quad (0,1)    \quad & \vline{} & \quad (1,0)    \quad & \vline{} & \quad (1,0)    \quad & \vline{}\\
\cline{2-9} \vline{} & \quad a^1_3 \ & \vline{} & \quad (0,1)    \quad & \vline{} & \quad (0,1)    \quad & \vline{} & \quad (1,1)    \quad & \vline{}\\
\cline{2-9}
\end{array}
\]
In the above game, $((\frac{1}{2},\ \frac{1}{2},\ 0),(\frac{1}{2},\ \frac{1}{2},\ 0))$ and $((0,\ 0,\ 1),(0,\ 0,\ 1))$ are the two Nash equilibrium strategies. Having started the algorithm from a random initial point, variation of the objective function value and the strategies are shown in the plots in Fig:\ref{fig:hm_ap} and Fig:\ref{fig:hm_obj}.

\subsection{A game with infinite Nash equilibria : }

\setlength{\arraycolsep}{0pt}
\[
\begin{array}{ccccccc}
\cline{2-6} \vline{} & \quad         & \vline{} & \quad a^2_1      \quad & \vline{} & \quad a^2_2      \quad & \vline{} \\
\cline{2-6} \vline{} & \quad a^1_1 \ & \vline{} & \quad (3,0)      \quad & \vline{} & \quad (12,0)     \quad & \vline{} \\
\cline{2-6} \vline{} & \quad a^1_2 \ & \vline{} & \quad (3,-2)     \quad & \vline{} & \quad (2,-5)     \quad & \vline{} \\
\cline{2-6}
\end{array}
\]

\begin{figure}[h]
 \begin{center}
 
 \begin{subfigure}[b]{0.46\textwidth}
  \includegraphics[width=\textwidth]{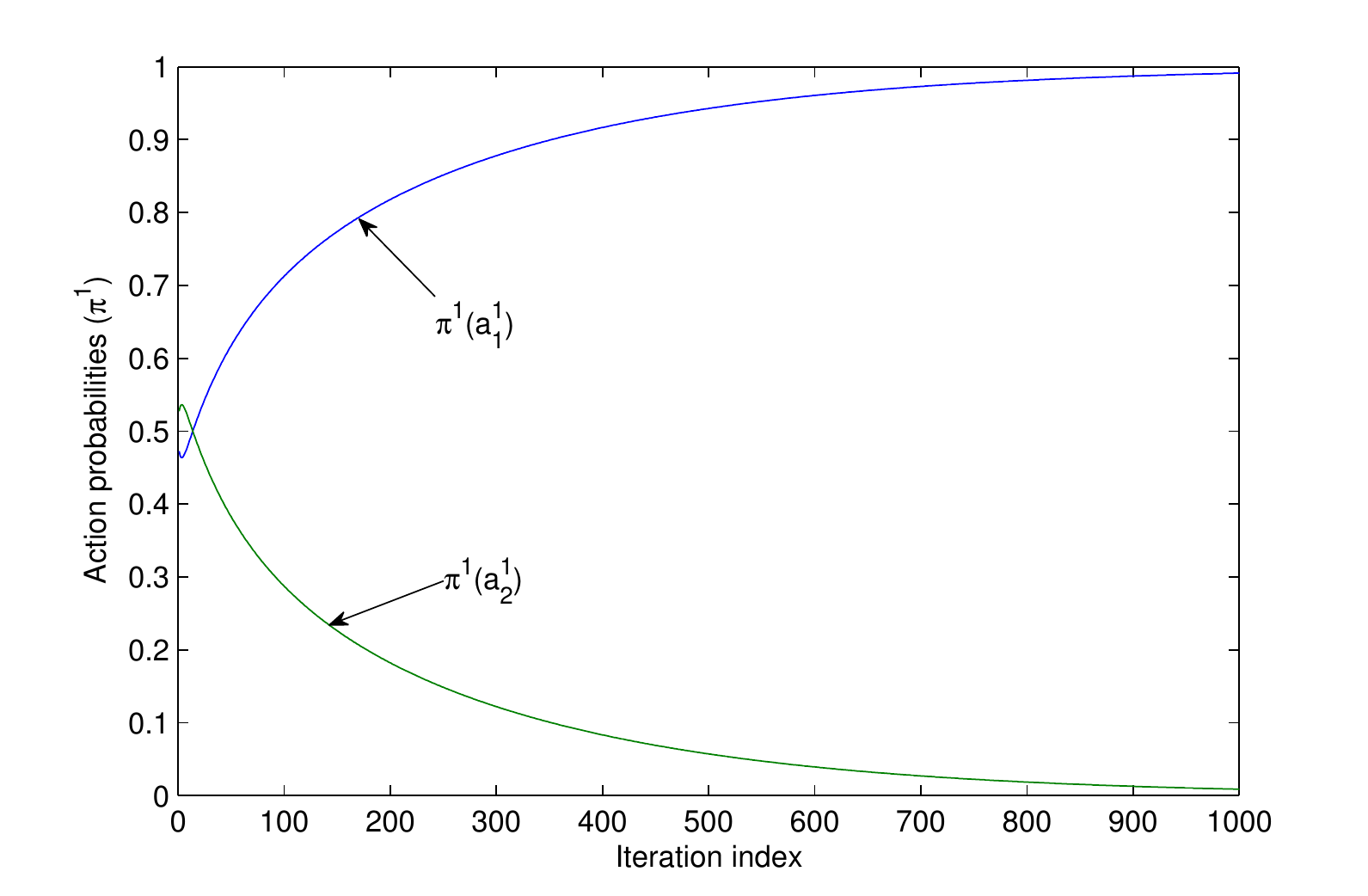}
  \caption{Action probabilities of player 1 vs iteration index.}
  \label{fig:ie_p1}
 \end{subfigure}
 \begin{subfigure}[b]{0.4\textwidth}
  \includegraphics[width=\textwidth]{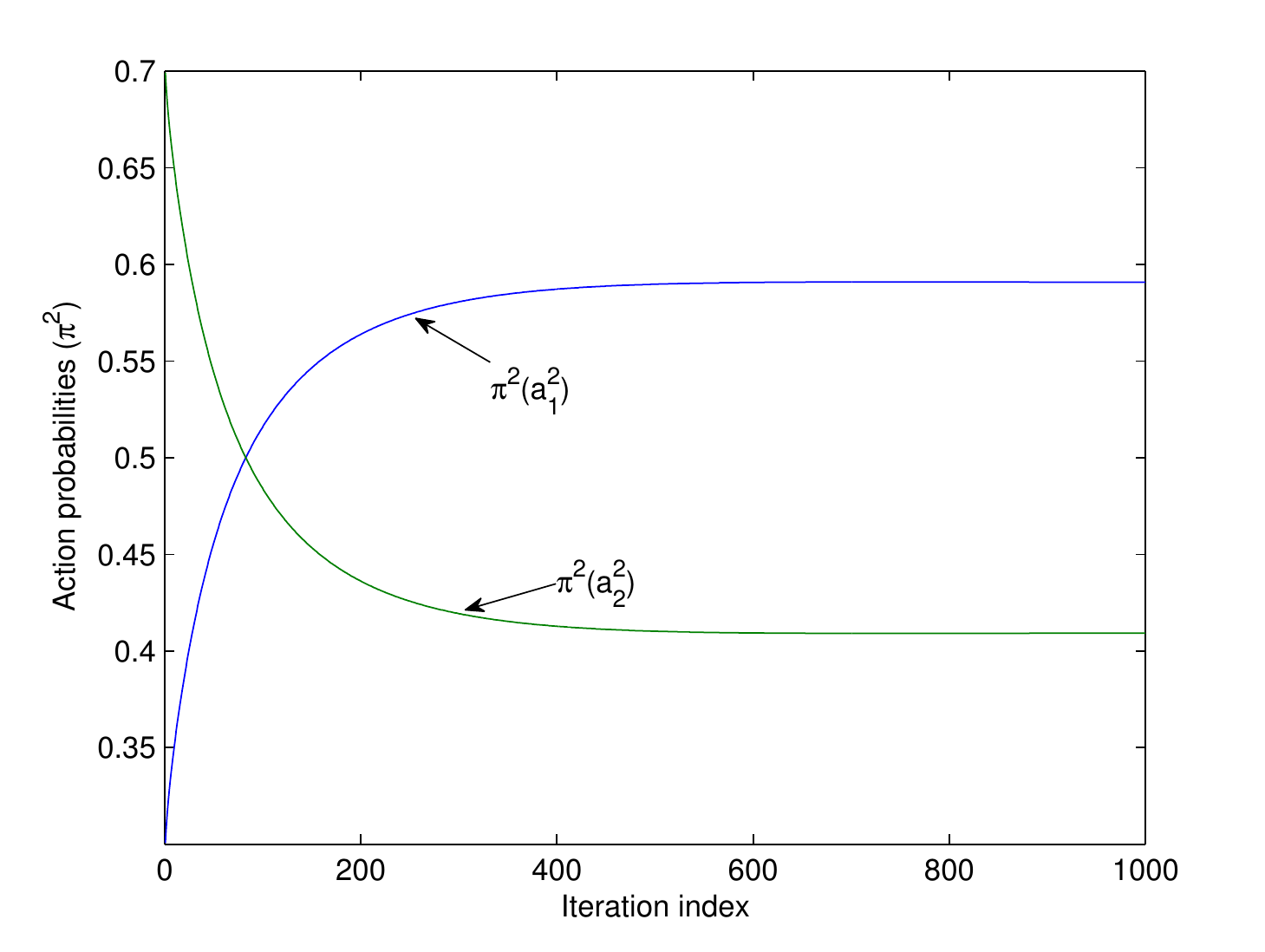}
  \caption{Action probabilities of player 2 vs iteration index.}
  \label{fig:ie_p2}
 \end{subfigure}
\caption{Action probabilities vs iteration index}
\label{fig:ie_ap}
\end{center}
\end{figure}

\begin{figure}[h]
\begin{center}
\scalebox{0.5}{\includegraphics{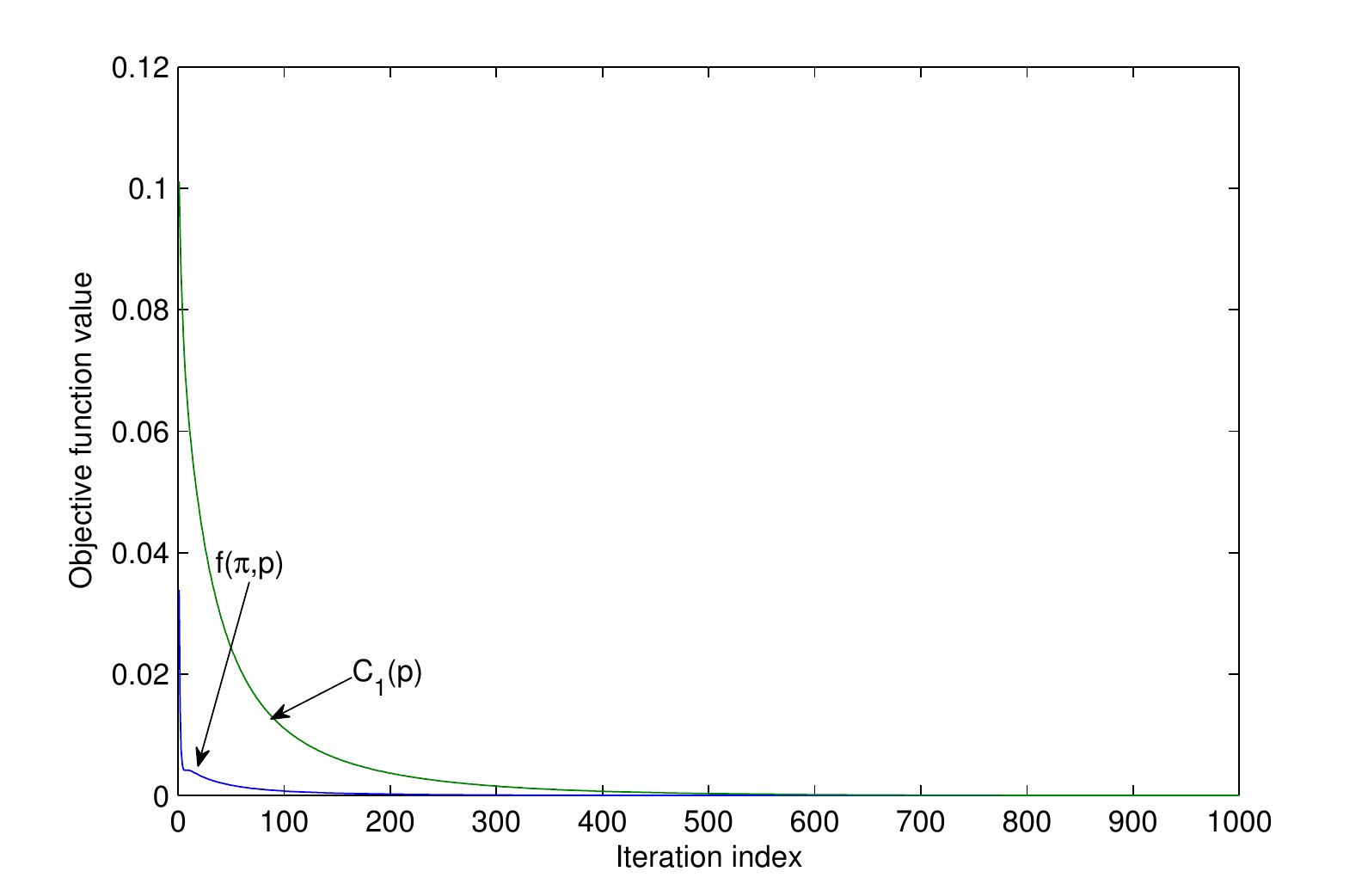}}
\caption{Objective function value vs iteration index.}
\label{fig:ie_obj}
\end{center}
\end{figure}

In the above game, $\{ ((\alpha,\ 1-\alpha),(1,\ 0)):0\leq \alpha \leq 1\} \cup \{ ((1,\ 0),(\alpha,\ 1-\alpha)):0\leq \alpha \leq 1\}$ is the set of Nash equilbria.  Having started the algorithm from a random initial point, variation of the objective function value and the strategies are shown in the plots in Fig:\ref{fig:ie_ap} and Fig:\ref{fig:ie_obj}. 

\newpage
\section{Summary and directions for future work.}
We have presented optimization problems ($O.P.1$ and $O.P.2$) such that the global minima of these optimization problems are Nash equilibria of the game $\Gamma$. The objective functions were shown to be bi-convex and in case of $O.P.1$ the objective function was also shown to be an invex function. We also considered a projected gradient descent scheme and proved that it converges to a partial optimum of the objective function. Even though the proof gaurantees convergence to the set of partial optimum in various test cases considered we have seen convergence to a Nash equilibrium strategy.

In future we wish to extend the above optimization problem formulation to discounted stochastic games and prove convergence to Nash equilibrium or construct a counter example where the algorithm converges to a partial optimum which is not a Nash equilibrium strategy.

\bibliographystyle{IEEEbib}

\end{document}